\documentclass[letterpaper,journal]{IEEEtran}
\usepackage[T1]{fontenc}
\usepackage{amsmath,amssymb,bm,mathtools}
\usepackage{cite}
\usepackage{booktabs,array,graphicx}
\usepackage{algorithm}
\usepackage[indLines=true,noEnd=false]{algpseudocodex}
\algrenewcommand{\algorithmicrequire}{\textbf{Input:}}
\algrenewcommand{\algorithmicensure}{\textbf{Output:}}
\algrenewcommand{\algorithmicindent}{1em}
\tikzset{algpxIndentLine/.style={draw=black,line width=0.35pt}}
\usepackage{tikz}
\usetikzlibrary{arrows.meta,positioning,calc,shapes.geometric}
\usepackage[hidelinks]{hyperref}
\hypersetup{
  pdftitle={Task-Oriented Network Ranging via Joint Graph and Schedule Design},
  pdfauthor={Zijian Zhang and Yuan Shen}
}
\usepackage{url}
\newcommand{\doi}[1]{\href{https://doi.org/#1}{{\urlstyle{same}\nolinkurl{#1}}}}
\usepackage{bm}

\DeclareMathAlphabet{\mathsfbr}{OT1}{cmss}{m}{n}%
\SetMathAlphabet{\mathsfbr}{bold}{OT1}{cmss}{bx}{n}%
\DeclareRobustCommand{\msf}[1]{%
  \ifcat\noexpand#1\relax\msfgreek{#1}\else\mathsfbr{#1}\fi%
}

\makeatletter
\newcommand{\msfgreek}[1]{\csname s\expandafter\@gobble\string#1\endcsname}
\makeatother

\DeclareFontEncoding{LGR}{}{} %
\DeclareSymbolFont{sfgreek}{LGR}{cmss}{m}{n}
\SetSymbolFont{sfgreek}{bold}{LGR}{cmss}{bx}{n}
\DeclareMathSymbol{\salpha}{\mathord}{sfgreek}{`a}
\DeclareMathSymbol{\sbeta}{\mathord}{sfgreek}{`b}
\DeclareMathSymbol{\sgamma}{\mathord}{sfgreek}{`g}
\DeclareMathSymbol{\sdelta}{\mathord}{sfgreek}{`d}
\DeclareMathSymbol{\sepsilon}{\mathord}{sfgreek}{`e}
\DeclareMathSymbol{\szeta}{\mathord}{sfgreek}{`z}
\DeclareMathSymbol{\seta}{\mathord}{sfgreek}{`h}
\DeclareMathSymbol{\stheta}{\mathord}{sfgreek}{`j}
\DeclareMathSymbol{\siota}{\mathord}{sfgreek}{`i}
\DeclareMathSymbol{\skappa}{\mathord}{sfgreek}{`k}
\DeclareMathSymbol{\slambda}{\mathord}{sfgreek}{`l}
\DeclareMathSymbol{\smu}{\mathord}{sfgreek}{`m}
\DeclareMathSymbol{\snu}{\mathord}{sfgreek}{`n}
\DeclareMathSymbol{\sxi}{\mathord}{sfgreek}{`x}
\DeclareMathSymbol{\somicron}{\mathord}{sfgreek}{`o}
\DeclareMathSymbol{\spi}{\mathord}{sfgreek}{`p}
\DeclareMathSymbol{\srho}{\mathord}{sfgreek}{`r}
\DeclareMathSymbol{\ssigma}{\mathord}{sfgreek}{`s}
\DeclareMathSymbol{\stau}{\mathord}{sfgreek}{`t}
\DeclareMathSymbol{\supsilon}{\mathord}{sfgreek}{`u}
\DeclareMathSymbol{\sphi}{\mathord}{sfgreek}{`f}
\DeclareMathSymbol{\schi}{\mathord}{sfgreek}{`q}
\DeclareMathSymbol{\spsi}{\mathord}{sfgreek}{`y}
\DeclareMathSymbol{\somega}{\mathord}{sfgreek}{`w}

\DeclareMathSymbol{\svarsigma}{\mathord}{sfgreek}{`c}

\DeclareMathSymbol{\sGamma}{\mathalpha}{sfgreek}{`G}
\DeclareMathSymbol{\sDelta}{\mathalpha}{sfgreek}{`D}
\DeclareMathSymbol{\sTheta}{\mathalpha}{sfgreek}{`J}
\DeclareMathSymbol{\sLambda}{\mathalpha}{sfgreek}{`L}
\DeclareMathSymbol{\sXi}{\mathalpha}{sfgreek}{`X}
\DeclareMathSymbol{\sPi}{\mathalpha}{sfgreek}{`P}
\DeclareMathSymbol{\sSigma}{\mathalpha}{sfgreek}{`S}
\DeclareMathSymbol{\sUpsilon}{\mathalpha}{sfgreek}{`U}
\DeclareMathSymbol{\sPhi}{\mathalpha}{sfgreek}{`F}
\DeclareMathSymbol{\sPsi}{\mathalpha}{sfgreek}{`Y}
\DeclareMathSymbol{\sOmega}{\mathalpha}{sfgreek}{`W}

\DeclareRobustCommand{\mcal}[1]{%
  \ifcat\noexpand#1\relax\mathnormal{#1}\else\cal{#1}\fi
}
\DeclareRobustCommand{\BM}[1]{%
  \ifcat\noexpand#1\relax\bm{\boldUppercaseItalicGreek{#1}}\else\bm{#1}\fi
}
\makeatletter
\newcommand{\boldUppercaseItalicGreek}[1]{\csname var\expandafter\@gobble\string#1\endcsname}
\makeatother
\newcommand{\rv}[1]{\MakeLowercase{\msf{#1}}}

\DeclareMathSizes{9}{9}{8}{8}
\newtheorem{theorem}{Theorem}
\newtheorem{proposition}{Proposition}
\newtheorem{lemma}{Lemma}

\newtheorem{remark}{Remark}
\newtheorem{definition}{Definition}
\makeatletter
\def\@begintheorem#1#2{\@IEEEtmpitemindent\itemindent\relax
  \topsep 0pt\normalfont\trivlist
  \item[]\textbf{\indent #1\ #2:}\ \itemindent\@IEEEtmpitemindent\relax}
\def\@opargbegintheorem#1#2#3{\@IEEEtmpitemindent\itemindent\relax
  \topsep 0pt\normalfont\trivlist
  \item[]\textbf{\indent #1\ #2:\ #3.}\ \itemindent\@IEEEtmpitemindent\relax}
\makeatother

\newcommand{\simroundwidth}{0.9871\columnwidth}
\begin{document}
\bstctlcite{IEEEBSTcontrol}
\title{Task-Oriented Network Ranging via Joint Graph and Schedule Design}
\author{Zijian~Zhang,~\IEEEmembership{Member,~IEEE}
and Yuan~Shen,~\IEEEmembership{Senior Member,~IEEE}%
\thanks{Z. Zhang and Y. Shen are with the Department of Electronic Engineering,
and Beijing National Research Center for Information Science and Technology,
Tsinghua University, Beijing 100084, China
(e-mail: zhangzij15@tsinghua.org.cn; shenyuan\_ee@tsinghua.edu.cn).
\textit{(Corresponding author: Yuan Shen.)}}}

\maketitle
\begin{abstract}
Measuring inter-node distances is the basis for network localization, whereas acquiring node positions does not require a fully connected graph of distances. Existing complete-ranging protocols for network localization require every node pair to exchange signals, which prolongs the ranging cycle. In this paper, we propose sparse signal-multiplexing network ranging (S-SM-NR), a task-oriented framework that adapts the transmit--listen schedule to the ranging task required by network localization. First, we establish the S-SM-NR framework by formulating the joint design of the task graph and the transmit--listen schedule. Then, we analyze the minimum round count required by S-SM-NR and establish a connectivity condition for network localization. Subsequently, we develop a closed-form joint design that constructs the sparse task graph and the balanced transmit--listen schedule. In particular, the joint design can achieve the minimum round count under the complete physical connectivity. Finally, simulations show that the proposed S-SM-NR scheme reduces the occupied-round count and improves the localization accuracy by shortening the ranging cycle.
\end{abstract}

\section{Introduction}\label{sec:introduction}
\IEEEPARstart{N}{etwork} localization estimates the positions of
multiple nodes to support cooperative sensing and multi-robot
operation~\cite{li2026scalable,zhao2025stereo}. In range-based
localization, inter-node distances provide geometric constraints, from
which the node positions can be recovered. Network ranging schemes can obtain
these distances through the wireless signal exchanges among nodes~\cite{zhao2022integrated,zhang2019efficient}.
In mobile scenarios, the inter-node distances should be updated frequently to track the
changes in the network geometry. However, as the network grows, the inter-node signal exchanges can occupy many transmission rounds, which decreases the update rate of node positions significantly. To support more frequent position updates, it is essential to reduce the number of occupied rounds so as to shorten the ranging cycle for distance acquisition.

To achieve this goal, concurrent ranging allows
multiple nodes to transmit signals in the same
round. Commercial half-duplex ultrawideband radios offer a low-cost way to implement
these concurrent transmissions~\cite{corbalan2020,schuh2024concurrent}. With such radio, however,
a node cannot receive while transmitting. To measure the distance of a required node pair,
the schedule needs to provide rounds in which one node transmits and the
other listens. Moreover, due to the limited signaling resources, receiver capacity limits the number of concurrent
signals that a listening node can resolve. Thus, concurrent network
ranging requires an efficient transmit--listen schedule that coordinates the inter-node signal
interactions and acquires all required distances within fewer rounds.

\subsection{Related Work}\label{sec:related}
Classical two-way ranging schemes estimate the distance between two nodes from
the transmission and reception times of request--reply signals~\cite{neirynck2016alternative}. In these protocols, the timestamp intervals can remove clock offsets, while clock-frequency deviations still affect the distance estimate. To address this issue, the alternative double-sided two-way ranging (AltDS-TWR) method uses three signal transmissions and suppresses the errors caused by clock-frequency deviations~\cite{neirynck2016alternative}.
Further studies examine the effects of signal timing and node movement
on ranging errors~\cite{shalaby2024timing,ferrero2026motion}.
However, repeating the signal exchanges independently for every node pair in networks
requires a large number of transmissions, which grows quadratically with the network size~\cite{neirynck2016alternative,zhang2022smnr}.

\begin{table*}[!t]
	\caption{Comparison of the primary ranging design objectives and capabilities.}
	\label{tab:related-comparison}
	\small
	\centering
	\setlength{\tabcolsep}{4pt}
	\begin{tabular*}{\textwidth}{@{\extracolsep{\fill}}lllll@{}}
		\toprule
		{\bf Capability} & \textbf{AltDS-TWR~\cite{neirynck2016alternative}} & {\bf SM-NR~\cite{zhang2022smnr}} & {\bf CC-SM-NR~\cite{ccsmnr2026}} & {\bf This work: S-SM-NR} \\
		\midrule
		Design input & Required node pairs & Complete graph & Complete graph & Localization requirement \\
		\midrule
		Task-graph selection & Prescribed & No selection & No selection & Jointly selected \\
		\midrule
		Primary ranging task & Pairwise ranging & Complete ranging & Complete ranging & Task-oriented ranging \\
		\midrule
		Schedule construction & Independent exchanges & Sequential broadcasts & Distinct codewords & Closed-form joint design \\
		\midrule
		Design scope & Pairwise ranging & Schedule only & Schedule only & Task graph and schedule \\
		\midrule
		Localization constraint & Not imposed & Not imposed & Not imposed & Global rigidity \\
		\bottomrule
	\end{tabular*}
\end{table*}

With the acquired distance measurements, network localization methods can
estimate the node positions by combining observations from multiple
node pairs~\cite{li2026scalable,li2024riemannian}.
For large networks, the calculations can be distributed among
nodes or organized into iterative position
updates~\cite{li2026scalable,wu2025finite,li2025uwbins}.
Some schemes are proposed to improve the localization accuracy by keeping position
estimates consistent and reducing the effect of unreliable
measurements~\cite{zhou2026distributed,yan2024robust}. Besides, robust optimization can also be utilized to reduce the effects of non-line-of-sight
propagation~\cite{long2024toa}.
When ranging observations are unreliable, measurements from other
sensors can provide additional position
constraints~\cite{wang2024factor}. In addition, existing studies have examined which measurements are needed to recover node
positions when clocks are independent or some observations are
missing~\cite{wen2023clock,cros2025pseudorange,chen2025relative}.
These studies address position estimation from available observations.
However, the number of rounds needed to obtain those observations still depends
on the signal-exchange schedule.

In network localization, reducing the measurement burden also involves selecting the nodes and the
links used for position acquisition. The node placement and selection should consider
position accuracy and communication
conditions~\cite{gomez2024deployment,li2025vehicles,lin2023}.
At the link level, the measured node pairs and exchanged observations
can be adapted to balance the position accuracy and communication
overhead~\cite{wang2025links}.
The measurement selection can be guided by information
measures~\cite{cai2026selection}, while system design can account for
localization accuracy and latency~\cite{petrucci2025multiobj}.
For the considered half-duplex network ranging, acquiring the selected measurements
requires feasible transmit--listen interactions under the receiver-capacity limit.

Ranging protocols address how these measurements are obtained through the inter-node
signal exchanges. Within the protocol, clock synchronization allows nodes to use timing
information from different local clocks, and has been studied for
cooperative localization and half-duplex
networks~\cite{gu2024clock,zino2025synchronization,lyu2025deferred}.
For concurrent transmissions, protocol frameworks coordinate radio
operations and time synchronization to reduce latency and energy
consumption~\cite{soprana2026samu}.
Concurrent reception can extract distance information from overlapping
signals, which allows multiple node pairs to be measured in the same
exchange~\cite{corbalan2020,schuh2024concurrent}.
Event-triggered exchanges of relative measurements and local estimates
reduce the communication burden~\cite{deng2026dual}.
Scheduling shared signal exchanges can improve the ranging efficiency in dense and dynamic
networks~\cite{hou2026swarm}, and periodic broadcasts can provide timestamps
for ranging across multiple node pairs~\cite{li2025onboard}.
Under half-duplex operation, to achieve inter-node ranging, these signal exchanges should be arranged
to provide observations in both directions for each required node pair~\cite{ccsmnr2026}.

To reduce the signaling burden in network ranging, some ranging schemes have been proposed. For example, the signal-multiplexing network ranging (SM-NR) scheme allows
each sequential broadcast to support ranging for multiple node pairs~\cite{zhang2022smnr}. In particular,
the broadcasts from a common reference
node provide a common timing interval to compensate for clock-frequency
deviations, while signal exchanges in both directions provide all
pairwise distances. To further reduce the localization latency, the concurrent coded SM-NR (CC-SM-NR) scheme is proposed for a half-duplex concurrent network. By dynamically assigning transmit and listen states across rounds, CC-SM-NR allows the concurrent transmissions among nodes to increase the update frequency of positions.~\cite{ccsmnr2026}.
Although both network ranging schemes improve ranging efficiency through schedule designs, they are designed for the complete-ranging task, which requires to measure all ranges of a given network~\cite{zhang2022smnr,ccsmnr2026}.

In fact, a suitable subset of inter-node distances
can provide sufficient geometric constraints for network localization
without requiring the complete distance graph~\cite{jackson2005,dokmanic2015}. Therefore, the complete-ranging requirement can impose more signal-exchange
constraints than network localization really needs, which inspires the idea of ranging with a sparse task graph. By jointly selecting the required
distances and designing the transmit--listen schedule, there is an opportunity to reduce the round count while preserving the geometric constraints required for network localization.

\subsection{Our Contributions}\label{sec:contributions}
To take advantage of this opportunity, this paper proposes sparse signal-multiplexing
network ranging (S-SM-NR) as a task-oriented framework.
The key idea is to jointly select the distances required for
localization and design the transmit--listen schedule to reduce the required
round count for network localization. Our contributions are summarized as follows.
\begin{enumerate}
\item We establish S-SM-NR as a general signal-processing framework for
the joint design of the task graph and the transmit--listen schedule.
By adapting the signal exchanges to the ranging task, the framework
can reduce the number of rounds required in network localization.
\item We determine the minimum round count and the distance requirements
for network localization within the S-SM-NR framework. Specifically, we derive the minimum round counts and lower bounds under different transmission constraints. A connectivity condition for network localization is established, which allows us to determine the minimum number of required distances.
\item We develop a closed-form joint design that constructs the sparse task graph and the balanced transmit--listen schedule. The proposed design balances the number of transmitters across rounds and assigns the two nodes joined by each task edge to different transmit rounds. Under complete physical connectivity, this schedule can achieve the minimum round count.
\end{enumerate}

In Table~\ref{tab:related-comparison}, we compare four typical network ranging schemes. As a baseline for pairwise ranging, AltDS-TWR is executed independently for each required node pair, without signal multiplexing across node pairs. The SM-NR scheme~\cite{zhang2022smnr} acquires all pairwise distances through sequential broadcasts, while the CC-SM-NR scheme~\cite{ccsmnr2026} can reduce the required round count by arranging the concurrent transmissions of half-duplex nodes.
Their designs focus on the transmit--listen schedule for complete ranging tasks.
In contrast, the proposed S-SM-NR jointly selects the task graph and the
schedule, which can shorten the ranging cycle while retaining sufficient distances for network localization.

\begin{table}[t]
	\caption{Graph notations used in this paper.}\label{tab:notation}
	\centering
	\small
	\begin{tabular}{@{}cl@{}}\toprule
		Symbol & Meaning\\\midrule
		$\mathcal V$ & Set of all vertices\\
		$\mathcal U$ & Set of non-reference vertices\\
		$\mathcal V_n$ & Vertex corresponding to node $n$\\
		$\mathcal E_{n_1,n_2}$ & Undirected edge between two vertices\\
		$\mathcal G_{\rm phy}$ & Physical connectivity graph\\
		$\mathcal H$ & Non-reference task graph\\
		$\mathcal F$ & Required graph for network localization\\
		$\mathcal F_{\rm av}(\bm X)$ & Available graph under schedule $\bm X$\\
		$\mathcal E_{\mathcal G}$ & Edge set of a graph $\mathcal G$\\
		\bottomrule\end{tabular}
\end{table}

\subsection{Organization and Notation}\label{sec:organization}
The remainder of this paper is organized as follows.
In Section~\ref{sec:model}, we present the network model and formulate the
joint graph and schedule design problem. Section~\ref{sec:analysis}
derives minimum round counts and lower bounds, and establishes the
connectivity condition for network localization.
In Section~\ref{sec:algorithms}, we develop the closed-form joint design
that achieves the minimum round count. In Section~\ref{sec:numerical}, we compare the round
counts and position accuracy under different system parameters. In Section~\ref{sec:conclusion},
the conclusions are drawn.

\textit{Notation:} Deterministic scalars, vectors, and matrices are
denoted by italic $x$, bold lowercase $\bm{x}$, and bold uppercase
$\bm{X}$ symbols, respectively. Graphs, vertices, and edges use calligraphic symbols
such as $\mathcal G$, $\mathcal V_n$, and $\mathcal E_{n_1,n_2}$,
respectively. Sets and set families also use calligraphic symbols such
as $\mathcal A$. Sans-serif symbols $\rv{x}$ and $\msf{X}$ denote
random quantities. The symbols $\|\cdot\|$ and
$\mathbb R$ denote the Euclidean norm and the set of real numbers,
respectively. For sets,
$|\mathcal A|$, $\mathcal A\setminus\mathcal B$, and
$\mathcal A\subseteq\mathcal B$ denote set size, set difference,
and inclusion, respectively. The operators $\lfloor\cdot\rfloor$ and
$\lceil\cdot\rceil$ denote floor and ceiling, and $\binom Rw$ denotes
a binomial coefficient. The indicator $\mathbb I\{\text{\it condition}\}$
equals one when the \emph{condition} holds and zero otherwise. The complete graph on $m$ vertices is
$\mathcal K_m$. For $m\ge3$, the cycle graph $\mathcal C_m$ connects the
vertices in order, with the last vertex connected back to the first.
The complete bipartite graph $\mathcal K_{a,b}$, which consists of disjoint
parts with $a$ and $b$ vertices, contains all edges between the parts
and no edges within either part. The \emph{graph join} $\vee$ connects every vertex
in one graph to every vertex in a disjoint second graph.

\section{System Model and Problem Formulation}\label{sec:model}
In this section, we establish the S-SM-NR framework by modeling the
ranging task and the transmit--listen schedule. In
Section~\ref{sec:network-clock}, we introduce the network capabilities
and assumptions. Section~\ref{sec:ranging-cycle} describes the signal
exchanges used to acquire the required distances. In
Section~\ref{sec:schedule-model}, we define the task graph and the
transmit--listen schedule. Section~\ref{sec:measurement-graphs}
distinguishes the required and available measurement graphs. Finally,
in Section~\ref{sec:joint-formulation}, we formulate the joint design
problem to minimize the round count in network localization.

\subsection{Network Capabilities and Assumptions}\label{sec:network-clock}
The network model specifies the geometry and communication constraints
for ranging. Consider a network of $N$ nodes represented by the vertex
set $\mathcal V=\{\mathcal V_0,\cdots,\mathcal V_M\}$, where $M=N-1$.
The indices $n,n_1,n_2$ identify nodes in $\{0,\cdots,M\}$. Node
$n$, represented by vertex $\mathcal V_n$, has position
$\bm p_n\in\mathbb R^2$. The distance between any two nodes is defined as
\begin{equation}
d_{n_1,n_2}=\|\bm p_{n_1}-\bm p_{n_2}\|.
\end{equation}
For each pair of distinct vertices, $d_{n_1,n_2}$ corresponds to the
undirected edge
$\mathcal E_{n_1,n_2}=\{\mathcal V_{n_1},\mathcal V_{n_2}\}$.
Vertex $\mathcal V_0$ represents the reference node, which supplies
a common timing interval to compensate for clock-frequency deviations
among nodes. All other
nodes are called non-reference nodes, and
$\mathcal U=\mathcal V\setminus\{\mathcal V_0\}$ contains their vertices.
For ease of reference, the graph notations are summarized in
Table~\ref{tab:notation}.

In practical systems, ranging feasibility depends on physical connectivity, half-duplex
operation, and receiver capacity~\cite{ccsmnr2026}. We define the \emph{physical
connectivity graph} $\mathcal G_{\rm phy}=(\mathcal V,\mathcal E_{\rm phy})$
to identify the node pairs with wireless links in both directions. Particularly, the reference node $\mathcal V_0$, as the first node to initiate ranging, has wireless links in both directions
to every non-reference node.\footnote{These links are necessary for the
reference-based ranging protocol, since each non-reference node should receive the
reference broadcasts for clock-frequency compensation. These links also enable ranging between the reference node and the non-reference nodes.} Each node uses an independent clock and cannot
receive while transmitting. Besides, finite orthogonal radio resources can limit the
number of concurrent signals that a listening node can resolve~\cite{ccsmnr2026}. To account for this limit, we
define the \emph{receiver capacity}, i.e., $K_{\max}\ge1$.

\subsection{Network Ranging Scheme}\label{sec:ranging-cycle}
\begin{figure}[!t]
\centering
\pgfdeclarelayer{rangingforeground}
\pgfsetlayers{main,rangingforeground}
\begin{tikzpicture}[x=1cm,y=1cm,
 every node/.style={font=\fontsize{9}{11}\selectfont},
 event/.style={circle,fill=black,inner sep=1.6pt},
 reception/.style={circle,draw=black,fill=white,inner sep=1.6pt},
 signal/.style={-{Stealth[length=2.2mm,width=1.5mm]},line width=1pt,
   shorten <=2.8pt,shorten >=2.8pt},
 reference/.style={signal,draw=blue},
 reciprocal/.style={signal,draw=red},
 intervalarrow/.style={-{Stealth[length=2.1mm,width=1.4mm]},
   draw=orange!55!black,line width=.9pt,shorten <=1pt},
 axishead/.style={-{Stealth[length=1.8mm,width=1.3mm]},
   draw=black!70,line width=.8pt}]
\path[use as bounding box](-.6,-.72) rectangle(8.05,3.57);
\foreach \yy/\name in {2.45/0,1.15/{n_1},-.15/{n_2}}{
 \draw[black!45,line width=.5pt] (.35,\yy)--(7.8,\yy);
 \node[anchor=east] at(.2,\yy){$\mathcal V_{\name}$};
}
\node[fill=white,inner sep=2pt](referenceinterval) at(3.9,3.14){Common reference interval $T$};
\node[align=center] at(1,2.85){Initial};
\node[align=center] at(6.8,2.85){Final};
\node[event] at(1,2.45){};
\node[event] at(6.8,2.45){};
\foreach \xx/\yy in {1.60/1.15,1.7/-.15,7.40/1.15,7.5/-.15}{
 \node[reception] at(\xx,\yy){};
}
\node[event] at(2.65,1.15){};
\node[event] at(4.95,-.15){};
\foreach \xx/\yy in {3.12/-.15,5.42/1.15,3.25/2.45,5.65/2.45}{
 \node[reception] at(\xx,\yy){};
}
\node[anchor=south] at(2.35,1.28){TX};
\node[anchor=south] at(5.74,1.28){RX};
\node[anchor=north] at(3.12,-.29){RX};
\node[anchor=north] at(4.95,-.29){TX};
\node[anchor=north] at(7.78,-.3){Time};
\foreach \yy in {2.45,1.15,-.15}{
 \node[fill=white,inner sep=1pt] at(3.9,\yy){$\cdots$};
}
\node[inner sep=0pt] at(0,1.80){$\vdots$};
\node[inner sep=0pt] at(0,.50){$\vdots$};
\node[inner sep=0pt] at(0,-.50){$\vdots$};
\begin{pgfonlayer}{rangingforeground}
 \draw[reference] (1,2.45)--(1.60,1.15);
 \draw[reference] (1,2.45)--(1.7,-.15);
 \draw[reference] (6.8,2.45)--(7.40,1.15);
 \draw[reference] (6.8,2.45)--(7.5,-.15);
 \draw[reciprocal] (2.65,1.15)--(3.12,-.15);
 \draw[reciprocal] (4.95,-.15)--(5.42,1.15);
 \draw[signal,black!75,dash pattern=on 2.5pt off 1.5pt]
   (2.65,1.15)--(3.25,2.45);
 \draw[signal,black!75,dash pattern=on 2.5pt off 1.5pt]
   (4.95,-.15)--(5.65,2.45);
 \draw[intervalarrow] (referenceinterval.west)--(1.0,3.14);
 \draw[intervalarrow] (referenceinterval.east)--(6.8,3.14);
 \foreach \yy in {2.45,1.15,-.15}{
  \draw[axishead] (7.6,\yy)--(7.8,\yy);
 }
\end{pgfonlayer}
\end{tikzpicture}
\caption{An illustration of the signal interactions for a required node pair. Node ${\cal V}_{n_1}$ and node ${\cal V}_{n_2}$ independently receive the two broadcasts from the reference node $\mathcal V_0$, which provide a common interval for clock-frequency compensation. The node pair exchanges reciprocal ranging signals.}
\label{fig:ranging-events}
\end{figure}
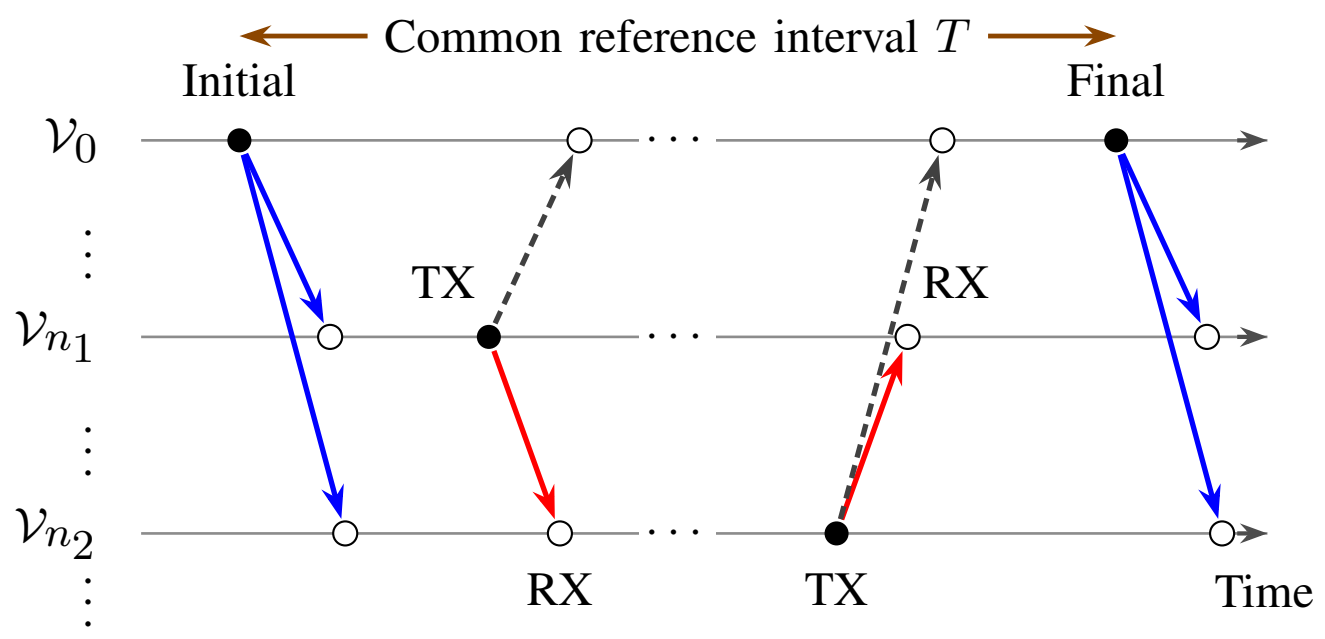

To acquire the task-required distances under asynchronous clocks, we combine
reciprocal signal exchanges with two reference broadcasts. This
reference-based ranging scheme compensates for clock-frequency
deviations among nodes~\cite{zhang2022smnr,zhao2024timeliness,ccsmnr2026}.
Let $R$ denote the intermediate-round count. TX and RX denote
transmission and reception events, respectively. In a ranging cycle,
each node records the timestamps of all TX and RX events in the
following three stages.
\begin{enumerate}
\item \emph{Initial reference broadcast:} The reference node $\mathcal V_0$ broadcasts, and all non-reference
nodes listen.
\item \emph{Intermediate transmit--listen rounds:} The non-reference nodes
execute $R$ transmit--listen rounds subject to the receiver capacity, while
the reference node listens. Each non-reference node transmits at least
once. To measure a pairwise distance, the two nodes need at least
one TX--RX interaction in each direction~\cite{ccsmnr2026}.
\item \emph{Final reference broadcast:} The reference node $\mathcal V_0$ broadcasts again, and all non-reference nodes listen.
\end{enumerate}

In Fig.~\ref{fig:ranging-events}, we illustrate the two reciprocal TX--RX
interactions used to recover $d_{n_1,n_2}$. The first transmitter measures
the interval from sending its signal to receiving the reply. The second
measures the interval from receiving that signal to sending its reply.
On a common time scale, the difference between these intervals
equals the sum of the forward and reverse propagation delays. To compensate for clock-frequency
deviations, the two broadcasts from $\mathcal V_0$ provide a common
reference interval $T$. Using the reference interval $T$, the nodes convert
timestamp intervals measured by their local clocks to a common time
scale before recovering the pairwise
distance, which effectively suppresses the meter-level ranging errors caused by clock asynchrony \cite{zhang2022smnr,zhao2024timeliness,ccsmnr2026}.

\subsection{Transmission Schedule}\label{sec:schedule-model}
To adapt the signal exchanges to the ranging task, we represent the
required distances by the \emph{task graph}
$\mathcal H=(\mathcal U,\mathcal E_{\mathcal H})$.
Each task edge in $\mathcal E_{\mathcal H}$ identifies a required
distance between two non-reference nodes, called the \emph{endpoints}
of the edge. An absent task edge means
that the distance is not required, regardless of whether a wireless
link exists. In ranging tasks, a feasible transmit--listen schedule should acquire every
required distance, assign at least one transmit round to every
non-reference node, and respect the receiver capacity.

For node index $1\le n\le M$ and intermediate-round index
$r\in\{1,\cdots,R\}$, let $x_{n,r}=1$ if $\mathcal V_n$ transmits
in round $r$, and let $x_{n,r}=0$ if it listens. The transmit--listen
schedule is represented by
$\bm X\in\{0,1\}^{M\times R}$, whose $(n,r)$-th element is $x_{n,r}$.
The $n$-th row of $\bm X$ is the transmit--listen \emph{codeword} for node ${\cal V}_n$ \cite{ccsmnr2026}.
Let $\mathcal R=\{1,\cdots,R\}$ denote the set of round indices.  The key schedule parameters are defined as
\begin{align}
	\mathcal A_n&=\{r\in\mathcal R:x_{n,r}=1\},\label{eq:support}\\
	w_n&=|\mathcal A_n|=\sum_{r=1}^{R}x_{n,r},\label{eq:weight}
\end{align}
where the
\emph{transmit support} $\mathcal A_n$ contains the rounds in which node
$n$ transmits, and $w_n$ denotes the number of transmit rounds of node $n$.
Each non-reference node transmits at least once, which implies that $w_n\ge1$
for $1\le n\le M$.

Furthermore, the transmit set $\mathcal S_r$ contains the nodes transmitting in round
$r$, and the round load $K_r$ is the number of these nodes, which are defined as
\begin{align}
 \mathcal S_r&=\{\mathcal V_n\in\mathcal U:x_{n,r}=1\},\label{eq:transmit-set}\\
 K_r&=|\mathcal S_r|=\sum_{n=1}^{M}x_{n,r}.\label{eq:round-load}
\end{align}
The round load is bounded by the receiver capacity, so we have $K_r\le K_{\max}$ in every round.

During the ranging stage, when
$\mathcal V_{n_1}$ transmits and $\mathcal V_{n_2}$ listens, the latter node
has a \emph{directed observation opportunity}. With this definition, the number of these
opportunities in one ranging cycle can be calculated as:
\begin{equation}
 \Gamma_{n_1\to n_2}(\bm X)
 =\sum_{r=1}^{R}x_{n_1,r}(1-x_{n_2,r}).\label{eq:gamma}
\end{equation}
The product $x_{n_1,r}(1-x_{n_2,r})$ is 1 when node $\mathcal V_{n_1}$ transmits and node $\mathcal V_{n_2}$ listens in round $r$, and is 0 otherwise.
To measure each required distance, the protocol requires {\it reciprocal coverage},
with at least one observation opportunity in each direction, leading to the constraint:
\begin{equation}
 \Gamma_{n_1\to n_2}\ge1,\quad\Gamma_{n_2\to n_1}\ge1,
 \qquad\mathcal E_{n_1,n_2}\in\mathcal E_{\mathcal H}.
 \label{eq:coverage}
\end{equation}
In this way, for every task edge, the constraint in \eqref{eq:coverage} allows the schedule to achieve the two directed observations
required by the network ranging protocol in Fig.~\ref{fig:ranging-events}.

\subsection{Measurement Graphs}\label{sec:measurement-graphs}
To distinguish the required distances from all acquired distances, we
define the required and available measurement graphs. In
Fig.~\ref{fig:ranging-events}, the reference broadcasts from
$\mathcal V_0$ and the return transmission from $\mathcal V_n$ provide
the distance between these nodes. We call the corresponding edges
\emph{reference edges} and collect them in
\begin{equation}
\mathcal E_0=\{\mathcal E_{0,n}:1\le n\le M\}.
\end{equation}
The required graph $\mathcal F$ contains the non-reference task edges
and all reference edges, which is defined as
\begin{equation}
 \mathcal F=\mathcal K_1\vee\mathcal H
   =(\mathcal V,\mathcal E_{\mathcal F}),\label{eq:cone}
\end{equation}
where $\mathcal E_{\mathcal F}=\mathcal E_0\cup\mathcal E_{\mathcal H}$ is the edge set of the graph $\cal F$, and $\mathcal K_1$ consists of the reference node $\mathcal V_0$.

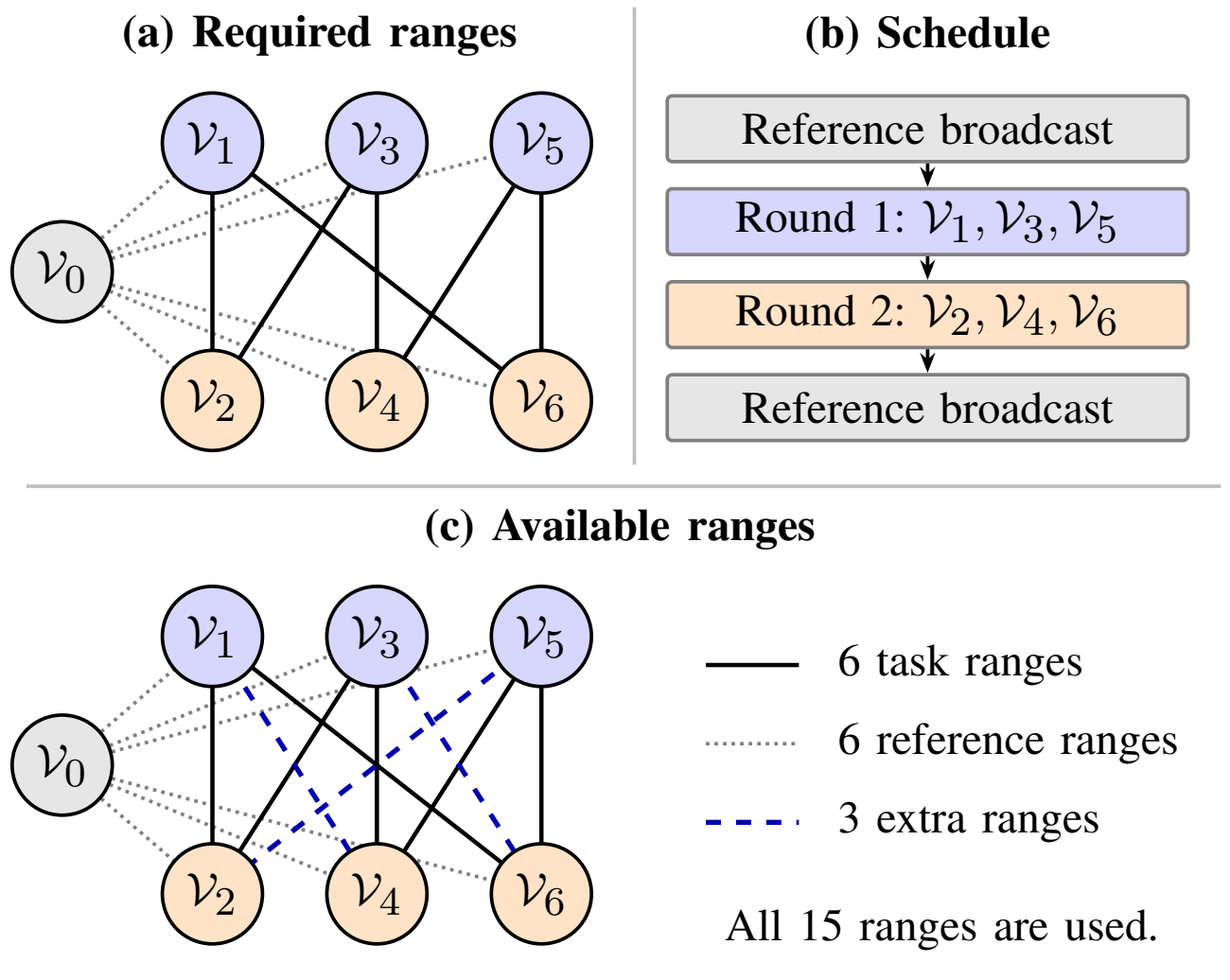
\begin{figure}[t]
\centering
\begin{tikzpicture}[x=1cm,y=1cm,
 every node/.style={font=\fontsize{9}{11}\selectfont},
 v/.style={circle,draw,line width=.7pt,minimum size=7mm,inner sep=.4pt},
 ga/.style={v,fill=blue!16},gb/.style={v,fill=orange!22},
 star/.style={black!50,densely dotted,line width=.7pt},
 cert/.style={black,line width=.85pt},
 extra/.style={blue!70!black,dashed,line width=1pt},
 slot/.style={draw=black!50,line width=.7pt,rounded corners=1pt,
   text width=3.5cm,minimum height=.46cm,align=center,inner sep=2pt}]
\path[use as bounding box](-.05,-.4) rectangle(8.55,6.55);
\foreach \tag/\off in {a/4.65,c/1.2}{
 \begin{scope}[yshift=\off cm]
 \coordinate (\tag0) at (.35,0);
 \foreach \n/\xx in {1/1.4,3/2.55,5/3.7}{\coordinate (\tag\n)at(\xx,.9);}
 \foreach \n/\xx in {2/1.4,4/2.55,6/3.7}{\coordinate (\tag\n)at(\xx,-.9);}
 \foreach \n in {1,...,6}{\draw[star](\tag0)--(\tag\n);}
 \draw[cert](\tag1)--(\tag2)--(\tag3)--(\tag4)--(\tag5)--(\tag6)--(\tag1);
 \ifnum\pdfstrcmp{\tag}{c}=0
  \foreach \na/\nb in {1/4,3/6,5/2}{\draw[extra](\tag\na)--(\tag\nb);}
 \fi
 \node[v,fill=black!10]at(\tag0){$\mathcal V_0$};
 \foreach \n in {1,3,5}{\node[ga]at(\tag\n){$\mathcal V_\n$};}
 \foreach \n in {2,4,6}{\node[gb]at(\tag\n){$\mathcal V_\n$};}
 \end{scope}}
\node[align=center,font=\fontsize{9}{11}\selectfont\bfseries]
 at(2.15,6.3){(a) Required ranges};
\node[align=center,font=\fontsize{9}{11}\selectfont\bfseries]
 at(6.4,6.3){(b) Schedule};
\draw[black!25,line width=.7pt](4.35,3.3)--(4.35,6.5);
\draw[black!25,line width=.7pt](.1,3.15)--(8.45,3.15);
\node[slot,fill=black!10](ref1)at(6.4,5.65){Reference broadcast};
\node[slot,fill=blue!16](round1)at(6.4,5.0)
 {Round 1: $\mathcal V_1,\mathcal V_3,\mathcal V_5$};
\node[slot,fill=orange!22](round2)at(6.4,4.35)
 {Round 2: $\mathcal V_2,\mathcal V_4,\mathcal V_6$};
\node[slot,fill=black!10](ref2)at(6.4,3.7){Reference broadcast};
\draw[-{Stealth[length=1.4mm]},line width=.7pt](ref1)--(round1);
\draw[-{Stealth[length=1.4mm]},line width=.7pt](round1)--(round2);
\draw[-{Stealth[length=1.4mm]},line width=.7pt](round2)--(ref2);
\node[align=center,font=\fontsize{9}{11}\selectfont\bfseries]
 at(4.25,2.85){(c) Available ranges};
\draw[cert](4.85,1.9)--(5.5,1.9);
\node[anchor=west]at(5.65,1.9){6 task ranges};
\draw[star](4.85,1.35)--(5.5,1.35);
\node[anchor=west]at(5.65,1.35){6 reference ranges};
\draw[extra](4.85,.8)--(5.5,.8);
\node[anchor=west]at(5.65,.8){3 extra ranges};
\node[align=center]at(6.5,.05){All 15 ranges are used.};
\end{tikzpicture}
\caption{Required and available ranges for 7 nodes
$\mathcal V_0,\cdots,\mathcal V_6$, with the reference node $\mathcal V_0$
and the receiver capacity $K_{\max}=3$.
The same schedule supplies the 3 extra ranges.}
\label{fig:concept}
\end{figure}
The schedule $\bm X$ can provide additional distances beyond the
required graph. We write the \emph{available graph} as
$\mathcal F_{\rm av}(\bm X)=(\mathcal V,\mathcal E_{\mathcal F_{\rm av}(\bm X)})$.
The edge set of $\mathcal F_{\rm av}(\bm X)$ contains all reference edges and every physically usable
non-reference edge whose endpoints have reciprocal observation
opportunities satisfying~\eqref{eq:coverage}. All distances in this graph
can be used for position estimation. A feasible schedule satisfies
\begin{equation}
 \mathcal E_{\mathcal F}\subseteq\mathcal E_{\mathcal F_{\rm av}(\bm X)}
       \subseteq\mathcal E_{\rm phy}.\label{eq:available-inclusion}
\end{equation}
For example, Fig.~\ref{fig:concept} illustrates how a schedule acquires both the required distances and additional available distances. It implies that, even if the distance measurement protocols are designed for a sparse task graph, it can still enable a large number of range estimates.

\subsection{Joint Graph and Schedule Design}\label{sec:joint-formulation}
The S-SM-NR framework adapts the transmit--listen schedule to the
required ranging task, with the objective of minimizing the
intermediate-round count. Including the
two reference broadcasts, the number of occupied rounds is
$
 L=R+2$.
We jointly select the required node pairs in the task graph
$\mathcal H$ and assign transmit rounds to the non-reference nodes in the
schedule $\bm X$. { To describe a feasible task graph for network localization, we provide the definition of \emph{generic global rigidity} as follows.}
\begingroup

\begin{definition}[Generic global rigidity]
A graph is \emph{generically globally rigid} in the plane if, for almost
all planar node configurations, the distances along its edges uniquely
determine the node positions up to a common translation, rotation, or
reflection of the entire network.
\end{definition}
\endgroup

Let $\mathcal G_{\rm loc}$ denote the class of graphs that are
generically globally rigid in the plane. To achieve the localization goal, one should restrict the choice of task edges to physical links, so
$\mathcal E_{\mathcal H}\subseteq\mathcal E_{\rm phy}$, as required by~\eqref{eq:available-inclusion}. Thus, the joint design problem can be formulated as
\begin{equation}
\begin{aligned}
 \underset{\mathcal H,\bm X,R}{\operatorname{minimize}}\quad &R\\
 \text{subject to}\quad
 &\mathcal K_1\vee\mathcal H\in\mathcal G_{\rm loc},\\
 &\Gamma_{n_1\to n_2},\Gamma_{n_2\to n_1}\ge1,
     \quad\mathcal E_{n_1,n_2}\in\mathcal E_{\mathcal H},\\
 &w_n\ge1,\quad 1\le n\le M,\\
 &K_r\le K_{\max},\quad 1\le r\le R,\\
 &x_{n,r}\in\{0,1\}.\label{eq:joint}
\end{aligned}
\end{equation}
The first constraint requires global rigidity of the required graph
$\mathcal F$. The next two constraints ensure reciprocal coverage for
every task edge and at least one transmission from every non-reference
node. The last two limit the round load $K_r$ to the receiver capacity
$K_{\max}$ and require binary entries $x_{n,r}$ in the schedule $\bm X$.

To solve this problem, the main difficulty is to select sufficient distances for localization
while reducing the round count \(R\). Concurrent transmissions can save
rounds, but nodes transmitting together cannot receive signals from one
another. Therefore, the task graph \(\mathcal H\) and the transmit--listen schedule
\(\bm X\) should be designed jointly under the receiver
capacity \(K_{\max}\). In particular, when the task graph $\mathcal H$ is prescribed, the schedule $\bm X$
is optimized subject to the signal-exchange constraints. This fixed-task
problem can provide a starting point for analyzing the joint graph and
schedule design.

\section{Performance Analysis}\label{sec:analysis}
In this section, we analyze the minimum round count and the distance
requirements for network localization within the S-SM-NR framework.\footnote{For the joint minimum-round analysis, we consider the ideal case with complete physical connectivity, i.e.,
		$\mathcal G_{\rm phy}=\mathcal K_N$, where only a subset of node pairs is selected for ranging.}

\subsection[Prescribed-Task Scheduling With Unlimited Receiver Capacity]{Prescribed-Task Scheduling With Unlimited Receiver Capacity}\label{sec:prescribed-task-analysis}
First, we determine the minimum round count for a prescribed
task graph $\mathcal H$ under unlimited receiver capacity.
We express reciprocal coverage through the transmit supports of
the schedule $\bm X$. Each endpoint needs to transmit in a round in which
the other listens. Thus, each support needs to contain a round absent
from the other support. We refer to this relation, in which neither
support contains the other, as \emph{incomparability}. We apply the
reciprocal-coverage condition in~\cite{ccsmnr2026} to each task edge,
giving the following proposition.

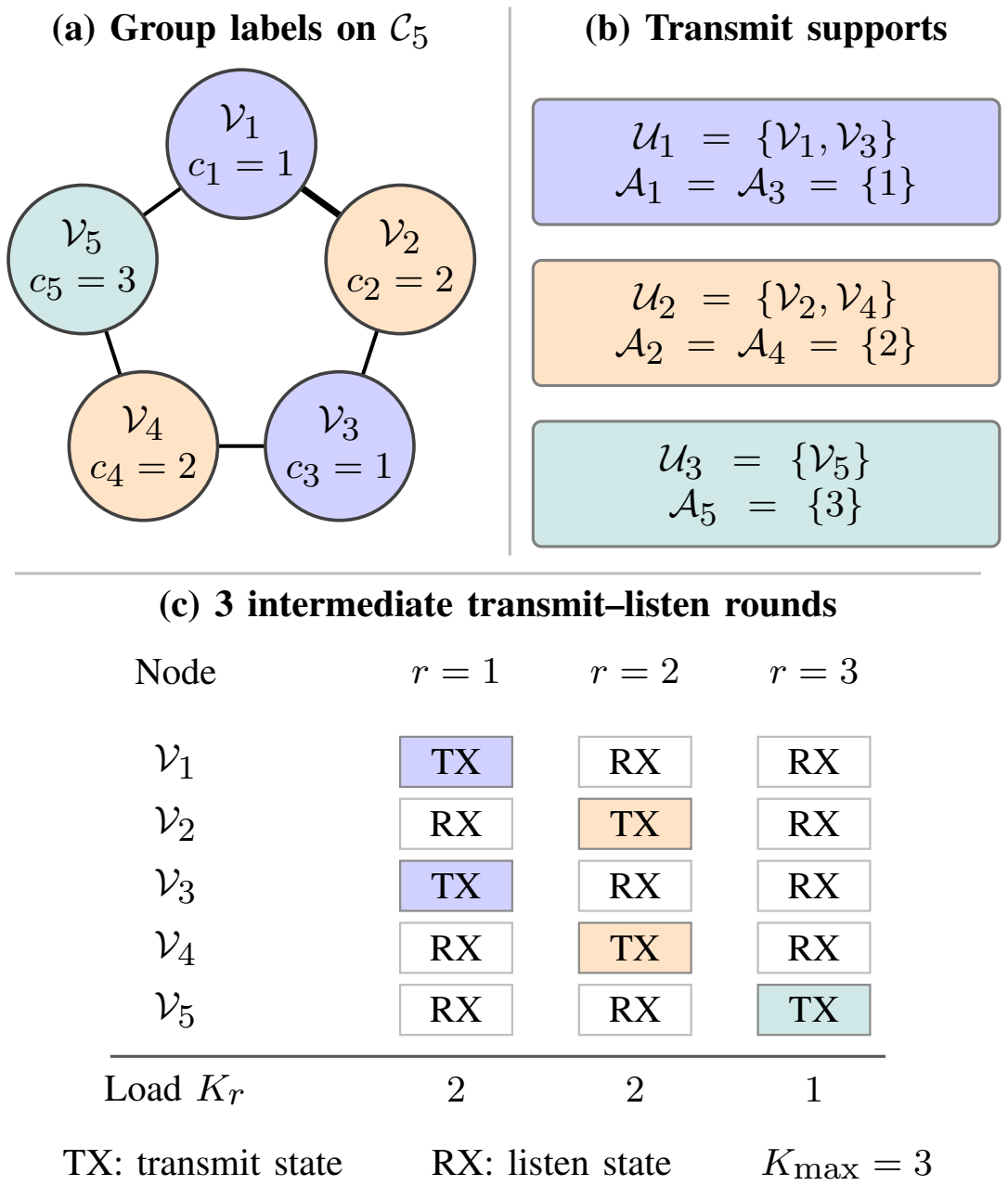
\begin{figure}[t]
\centering
\scalebox{0.85}{%
\begin{tikzpicture}[x=1cm,y=1cm,
 every node/.style={font=\fontsize{9}{11}\selectfont},
 v/.style={circle,draw=black!75,line width=.8pt,minimum size=10.5mm,
   inner sep=1.2pt,align=center},
 card/.style={draw=black!50,line width=.7pt,rounded corners=2pt,
   text width=3.7cm,minimum height=1.05cm,align=center,inner sep=3pt}]
\path[use as bounding box](0,0)rectangle(8.5,10);
\draw[black!25,line width=.7pt](4.35,5.35)--(4.35,9.95);
\draw[black!25,line width=.7pt](.2,5.2)--(8.3,5.2);
\node[font=\fontsize{9}{11}\selectfont\bfseries]at(2.1,9.75)
 {(a) Group labels on $\mathcal C_5$};
\node[font=\fontsize{9}{11}\selectfont\bfseries]at(6.5,9.75)
 {(b) Transmit supports};
\begin{scope}[shift={(2.1,7.4)}]
 \foreach \n in {1,...,5}{
  \coordinate(p\n)at({90-72*(\n-1)}:1.4);}
 \draw[line width=.9pt](p1)--(p2)--(p3)--(p4)--(p5)--(p1);
 \draw[line width=1.6pt](p1)--(p2);
 \foreach \n/\colorid/\shade in
  {1/1/blue!18,2/2/orange!22,3/1/blue!18,4/2/orange!22,5/3/teal!18}{
  \node[v,fill=\shade]at(p\n){$\mathcal V_{\n}$\\$c_{\n}=\colorid$};}
\end{scope}
\node[card,fill=blue!18]at(6.5,8.65)
 {$\mathcal U_1=\{\mathcal V_1,\mathcal V_3\}$\\
  $\mathcal A_1=\mathcal A_3=\{1\}$};
\node[card,fill=orange!22]at(6.5,7.3)
 {$\mathcal U_2=\{\mathcal V_2,\mathcal V_4\}$\\
  $\mathcal A_2=\mathcal A_4=\{2\}$};
\node[card,fill=teal!18]at(6.5,5.95)
 {$\mathcal U_3=\{\mathcal V_5\}$\\$\mathcal A_5=\{3\}$};
\node[font=\fontsize{9}{11}\selectfont\bfseries]at(4.25,4.92)
 {(c) 3 intermediate transmit--listen rounds};
\node at(1.55,4.38){Node};
\foreach \r in {1,2,3}{\node at({2.4+1.5*\r},4.38){$r=\r$};}
\foreach \n/\colorid/\shade in
 {1/1/blue!18,2/2/orange!22,3/1/blue!18,4/2/orange!22,5/3/teal!18}{
 \node at(1.55,{4.14-.52*\n}){$\mathcal V_{\n}$};
 \foreach \r in {1,2,3}{
  \ifnum\r=\colorid
   \filldraw[fill=\shade,draw=black!45,line width=.5pt]
    ({2.4+1.5*\r-.47},{4.14-.52*\n-.21})
    rectangle({2.4+1.5*\r+.47},{4.14-.52*\n+.21});
   \node at({2.4+1.5*\r},{4.14-.52*\n}){TX};
  \else
   \draw[black!25,line width=.5pt]
    ({2.4+1.5*\r-.47},{4.14-.52*\n-.21})
    rectangle({2.4+1.5*\r+.47},{4.14-.52*\n+.21});
   \node at({2.4+1.5*\r},{4.14-.52*\n}){RX};
  \fi
 }}
\draw[black!65,line width=.7pt](1,1.15)--(7.5,1.15);
\node at(1.55,.87){Load $K_r$};
\foreach \r/\load in {1/2,2/2,3/1}{
 \node at({2.4+1.5*\r},.87){$\load$};}
\node at(4.25,.25){TX: transmit state \qquad RX: listen state \qquad $K_{\max}=3$};
\end{tikzpicture}
}
\caption{A 3-round schedule for the 5-vertex cycle graph $\mathcal C_5$
with $N=6$ and $K_{\max}=3$. Colors mark groups $\mathcal U_\ell$, where
node $\mathcal V_n$ has label $c_n=\ell$ and transmit support
$\mathcal A_n=\{\ell\}$. TX and RX denote the transmit and listen states with
$x_{n,r}=1$ and $x_{n,r}=0$, respectively.}
\label{fig:coloring}
\end{figure}

\begin{proposition}[Incomparability condition for reciprocal coverage]\label{prop:reciprocal}
For a task edge
$\mathcal E_{n_1,n_2}=\{\mathcal V_{n_1},\mathcal V_{n_2}\}$,
reciprocal coverage in~\eqref{eq:coverage} holds if and only if
\begin{equation}
 \mathcal A_{n_1}\nsubseteq\mathcal A_{n_2},
 \qquad \mathcal A_{n_2}\nsubseteq\mathcal A_{n_1}.
 \label{eq:incomparable}
\end{equation}
\end{proposition}
\begin{IEEEproof}
For each direction, an observation opportunity occurs in a round
belonging to the support of the transmitting node but not that of the
listening node. Thus, the opportunity counts are
\begin{equation}
 \begin{aligned}
 \Gamma_{n_1\to n_2}(\bm X)
   &=|\mathcal A_{n_1}\setminus\mathcal A_{n_2}|,\\
 \Gamma_{n_2\to n_1}(\bm X)
   &=|\mathcal A_{n_2}\setminus\mathcal A_{n_1}|.
 \end{aligned}
 \label{eq:support-difference}
\end{equation}
Both counts are positive if and only if the supports are incomparable,
which establishes reciprocal coverage in~\eqref{eq:coverage}.
\end{IEEEproof}

Two nodes that use the same transmit--listen codeword always transmit
in the same rounds and listen in the same rounds. Neither node can receive
a ranging signal from the other. In Fig.~\ref{fig:coloring}, for example,
node $\mathcal V_1$ and node $\mathcal V_3$ can share a codeword because
no task edge joins them. This inspires us to group the nodes so that no task edge
lies within a group and use a color to label each group. Let $C\ge1$ denote the number of groups and
$c_n\in\{1,\cdots,C\}$ the label of $\mathcal V_n$. Nodes with label
$\ell\in\{1,\cdots,C\}$ form the group
$\mathcal U_\ell=\{\mathcal V_n\in\mathcal U:c_n=\ell\}$. A
\emph{proper coloring} assigns different labels to the endpoints of
every task edge. The \emph{chromatic number} $\chi(\mathcal H)$ is the
smallest number of groups in a proper coloring.\footnote{An optimal grouping exists for every finite task graph
and can be found by exhaustive search. Computing the minimum group count
$\chi(\mathcal H)$ is NP-hard, and exact algorithms may require exponential
time in the worst case.}

We assign one transmit--listen codeword to each group and choose
the transmit supports of these codewords to be pairwise incomparable.
By Proposition~\ref{prop:reciprocal}, each endpoint of a task edge then
has a round in which it transmits while the other listens. To relate the group count $C$ to the number of
rounds, we define
\begin{equation}
 \rho(C)=\min\left\{R\in\mathbb Z_{\ge0}:
        \binom{R}{\left\lfloor\frac{R}{2}\right\rfloor}\ge C\right\}.
 \label{eq:rho}
\end{equation}
The binomial coefficient counts the distinct $R$-round codewords with
exactly $\lfloor R/2\rfloor$ transmit rounds.

Let $R^\star_\infty(\mathcal H)$ denote the minimum intermediate-round
count under unlimited receiver capacity when only reciprocal task-edge
coverage is required. The following theorem determines this minimum
round count from the chromatic number $\chi(\mathcal H)$.
\begin{theorem}[Minimum rounds for a prescribed task graph]\label{thm:fixed}
Let $\mathcal H=(\mathcal U,\mathcal E_{\mathcal H})$ be a finite
simple graph with $M\ge1$ vertices.
If only reciprocal coverage of task edges is required, the minimum
intermediate-round count is
\begin{equation}
 R^\star_\infty(\mathcal H)=\rho\bigl(\chi(\mathcal H)\bigr).
 \label{eq:fixed-optimum}
\end{equation}
Requiring every non-reference node to transmit at least once gives the minimum
$\max\{1,\rho(\chi(\mathcal H))\}$.
\end{theorem}
\begin{IEEEproof}
See Appendix~\ref{app:fixed}.
\end{IEEEproof}
\begin{remark}
For complete ranging, $\mathcal H=\mathcal K_M$ and
$\chi(\mathcal H)=M$, so Theorem~\ref{thm:fixed} recovers the minimum
round count $\rho(M)$ in~\cite{ccsmnr2026}.
With unlimited receiver capacity, 6 groups can be scheduled in 4
	intermediate rounds by assigning one of the following codewords to each group:
	\begin{equation*}
		\mathtt{1100},\quad\mathtt{1010},\quad\mathtt{1001},\quad\mathtt{0110},\quad\mathtt{0101},\quad\mathtt{0011}.
	\end{equation*}
	The entries 1 and 0 denote transmission and listening, respectively. For example,
	nodes using codeword $\mathtt{1100}$ transmit in rounds 1 and 2 and
	listen in rounds 3 and 4. For any two groups, each has a transmit round
	in which the other listens. As a result, these codewords provide
	reciprocal coverage for all task edges between groups.
\end{remark}

\subsection[Prescribed-Task Scheduling With Finite Receiver Capacity]{Prescribed-Task Scheduling With Finite Receiver Capacity}\label{sec:finite-capacity-analysis}
Then, we account for finite receiver capacity while keeping the task graph fixed. We derive
lower bounds on the minimum feasible intermediate-round count $R^\star_{K_{\max}}(\mathcal H)$ for
a given $K_{\max}$. Counting at least one transmission per non-reference
node and at most $K_{\max}$ transmitters per round, together with the
unlimited-capacity bound $\rho(\chi(\mathcal H))$ from
Theorem~\ref{thm:fixed}, gives~\eqref{eq:bounds-basic}. Reciprocal
coverage of each task edge requires two distinct transmit--listen rounds,
which gives~\eqref{eq:bounds-halfduplex}. Moreover, node $\mathcal V_n$
needs to receive a transmission from each node connected to it by a task
edge. These nodes are the task neighbors of $\mathcal V_n$, and their
number $D_n$ is the degree of $\mathcal V_n$ in $\mathcal H$. The receptions
take place in the listening rounds of $\mathcal V_n$, since a half-duplex
node cannot receive while transmitting. For $D_n>0$, counting these listening rounds together with
one transmit round gives~\eqref{eq:bounds-degree}.

We further count the required observations within a node subset to
account for how task edges are distributed. For
$\mathcal W\subseteq\mathcal U$, the induced graph
$\mathcal H[\mathcal W]$ contains the vertices in $\mathcal W$ and all
task edges between them, with the edge set
$\mathcal E_{\mathcal H[\mathcal W]}$. In each round, the transmitting
and listening nodes divide $\mathcal W$ into two groups. This division
defines a \emph{cut}, and each task edge joining the groups can provide
one directed observation opportunity in that round. When the induced edge set is
nonempty, let $b_{K_{\max}}(\mathcal W)$ denote the maximum number of
these crossing edges over all choices of at most $K_{\max}$ transmitters
in $\mathcal W$. This quantity limits the number of these opportunities
available in one round within $\mathcal W$. Since each task edge
needs observation opportunities in both directions, this limit
yields~\eqref{eq:bounds-cut}.

The following proposition combines these
bounds, while the formal definition of $b_{K_{\max}}(\mathcal W)$ is given
in Appendix~\ref{app:bounds}.

\begin{proposition}[Bounds for a prescribed task graph]\label{prop:finite}
For a prescribed task graph $\mathcal H$, the minimum round count
satisfies
\begin{align}
 R^\star_{K_{\max}}(\mathcal H)&\ge
 \max\left\{\rho(\chi(\mathcal H)),\left\lceil\frac{M}{K_{\max}}\right\rceil\right\},
 \label{eq:bounds-basic}\\
 R^\star_{K_{\max}}(\mathcal H)&\ge2
 \quad\text{if }\mathcal E_{\mathcal H}\ne\varnothing,
 \label{eq:bounds-halfduplex}\\
 R^\star_{K_{\max}}(\mathcal H)&\ge1+
 \left\lceil\frac{D_n}{K_{\max}}\right\rceil,
 \label{eq:bounds-degree}\\
 R^\star_{K_{\max}}(\mathcal H)&\ge
 \left\lceil\frac{2|\mathcal E_{\mathcal H[\mathcal W]}|}
 {b_{K_{\max}}(\mathcal W)}\right\rceil.
 \label{eq:bounds-cut}
\end{align}
where~\eqref{eq:bounds-degree} applies to every node with $D_n>0$.
The bound in~\eqref{eq:bounds-cut} applies to every subset
$\mathcal W\subseteq\mathcal U$ with
$\mathcal E_{\mathcal H[\mathcal W]}\ne\varnothing$.
\end{proposition}
\begin{IEEEproof}
See Appendix~\ref{app:bounds}.
\end{IEEEproof}

\begingroup

\begin{remark}
	Consider $N=6$, $M=5$, and $K_{\max}=3$.
	In Fig.~\ref{fig:coloring}, the 5 non-reference nodes form 3
	groups of sizes 2, 2, and 1. Each group transmits in a separate
	intermediate round. The resulting round loads $(2,2,1)$ satisfy the
	receiver-capacity limit. Theorem~\ref{thm:fixed} rules out any schedule
	with fewer than 3 intermediate rounds for this task graph.
	Therefore, the schedule is optimal and uses $L=5$ occupied rounds per ranging cycle.
\end{remark}
\endgroup

\begin{remark}
Proposition~\ref{prop:finite} shows how the receiver capacity and the task
edges can increase the required round count beyond
$\rho(\chi(\mathcal H))$. The load bound counts the required
transmissions, the neighbor bound counts one transmit round and the listening rounds
a node needs to receive a signal from each of its task neighbors, and the cut bound counts observations across vertex groups.
For example, at $K_{\max}=3$, $\mathcal K_{2,7}$
has chromatic number 2 and requires 4 rounds,
matching the degree bound in~\eqref{eq:bounds-degree}.
For a feasible schedule, the part with 2 nodes transmits in one
round. The remaining 7 nodes transmit in three further rounds, giving
round loads $(2,3,3,1)$.
The cycle graph $\mathcal C_9$ has chromatic number 3 and
requires 3, matching the load bound in~\eqref{eq:bounds-basic}.
We achieve these 3 rounds by assigning the indices 1, 2, and 3
repeatedly along the cycle, with round loads $(3,3,3)$. Thus, under finite
receiver capacity, a smaller chromatic number
$\chi(\mathcal H)$ need not give a shorter schedule.
\end{remark}

\subsection[Joint Minimum-Round Analysis for Network Localization]{Joint Minimum-Round Analysis for Network Localization}\label{sec:joint-round-analysis}\label{sec:localization-topology}

We now determine the minimum round count over both task graphs and schedules. First, we identify task graphs whose required graph $\mathcal F=\mathcal K_1\vee\mathcal H$ is generically globally rigid, so that the selected distances support network localization.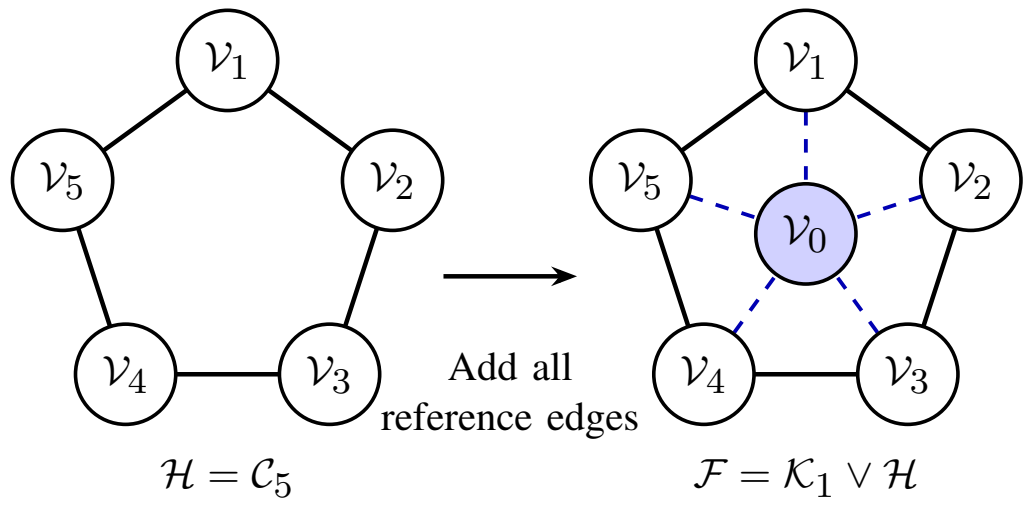
\begin{figure}[t]
\centering
\scalebox{0.85}{%
\begin{tikzpicture}[x=1cm,y=1cm,
 every node/.style={font=\fontsize{9}{11}\selectfont},
 v/.style={circle,draw,line width=.8pt,fill=white,
   minimum size=7.8mm,inner sep=1.3pt}]
\path[use as bounding box](0,.68)rectangle(8.5,4.70);
\foreach \offset/\tag in {2/a,6.5/b}{
 \begin{scope}[shift={(\offset,2.88)}]
  \foreach \n in {1,...,5}{
   \coordinate(\tag\n)at({90-72*(\n-1)}:1.35);}
  \ifdim\offset pt>3pt
   \foreach \n in {1,...,5}{
    \draw[blue!70!black,dashed,line width=.85pt](0,0)--(\tag\n);}
   \node[v,fill=blue!18]at(0,0){$\mathcal V_0$};
  \fi
  \draw[line width=1pt]
   (\tag1)--(\tag2)--(\tag3)--(\tag4)--(\tag5)--(\tag1);
  \foreach \n in {1,...,5}{
   \node[v]at(\tag\n){$\mathcal V_{\n}$};}
 \end{scope}
}
\node at(2,.98){$\mathcal H=\mathcal C_5$};
\node at(6.5,.98){$\mathcal F=\mathcal K_1\vee\mathcal H$};
\draw[-{Stealth[length=2mm]},line width=1pt](3.68,2.55)--(4.72,2.55);
\node[align=center]at(4.2,1.62){Add all\\reference edges};
\end{tikzpicture}
}
\caption{Required graph used in Lemma~\ref{lem:cone}. Adding the ranges from the reference node $\mathcal V_0$ to every non-reference node converts the task graph $\mathcal H$ into the required graph $\mathcal F$.}
\label{fig:cone}
\end{figure}
\subsubsection[Connectivity Condition for Network Localization]{Connectivity Condition for Network Localization}

As shown in Fig.~\ref{fig:cone}, $\mathcal F$ combines the task edges
with all reference edges between $\mathcal V_0$ and the non-reference
nodes. We establish the global-rigidity condition through the
connectivity of $\mathcal H$. Specifically, the task graph needs to be
connected and remain connected after deleting any non-reference vertex.
The graph $\mathcal H-\mathcal V_n$ is obtained by removing vertex
$\mathcal V_n$ and all edges connected to it. These two connectivity
conditions define \emph{2-vertex-connectivity}. The following lemma
shows that this connectivity test determines generic global rigidity
of $\mathcal F$ without using node coordinates.

\begin{lemma}[Global rigidity of the required graph]\label{lem:cone}
For $M\ge3$, $\mathcal F=\mathcal K_1\vee\mathcal H$ is generically
globally rigid in $\mathbb R^2$ if and only if $\mathcal H$ is
2-vertex-connected.
\end{lemma}
\begin{IEEEproof}
See Appendix~\ref{app:cone}.
\end{IEEEproof}
\begin{remark}
Using this connectivity condition, we can select a sparse task graph
that supports network localization. With all reference edges included, it is sufficient for the
task graph $\mathcal H$ to remain connected after any single
non-reference node is removed. Simple connectivity alone is insufficient,
since different parts of $\mathcal H$ may rely on a single node to
remain connected. In particular, a cycle graph through all $M$
non-reference nodes satisfies this condition with only $M$ task edges.
This provides a sparse task graph for the joint design of the required
distances and the transmit--listen schedule.
\end{remark}

\subsubsection[Joint Minimum Round Count]{Joint Minimum Round Count}
With this localization condition, we consider minimizing the round count over both task graphs and schedules. Let
$R^\star_{\rm joint}(N,K_{\max})$ denote the minimum intermediate-round
count over task graphs whose required graphs are generically globally
rigid and schedules satisfying the transmit--listen scheduling
constraints. The following theorem determines this minimum round count.

\begin{theorem}[Minimum round count for joint graph and schedule design]\label{thm:joint}
Let $N=M+1\ge5$ and let $K_{\max}\ge1$ be an integer.
Under complete physical connectivity and the scheduling constraints in
Section~\ref{sec:schedule-model}, the minimum round count is
\begin{equation}
 R^\star_{\rm joint}(N,K_{\max})
 =\max\left\{2,\left\lceil\frac{N-1}{K_{\max}}\right\rceil\right\}.
 \label{eq:joint-optimum}
\end{equation}
This minimum round count can be achieved with each non-reference node transmitting in only one intermediate round. The ranging cycle uses $N+1$ individual broadcasts and occupies
$R^\star_{\rm joint}+2$ rounds.
\end{theorem}
\begin{IEEEproof}
See Appendix~\ref{app:joint}.
\end{IEEEproof}

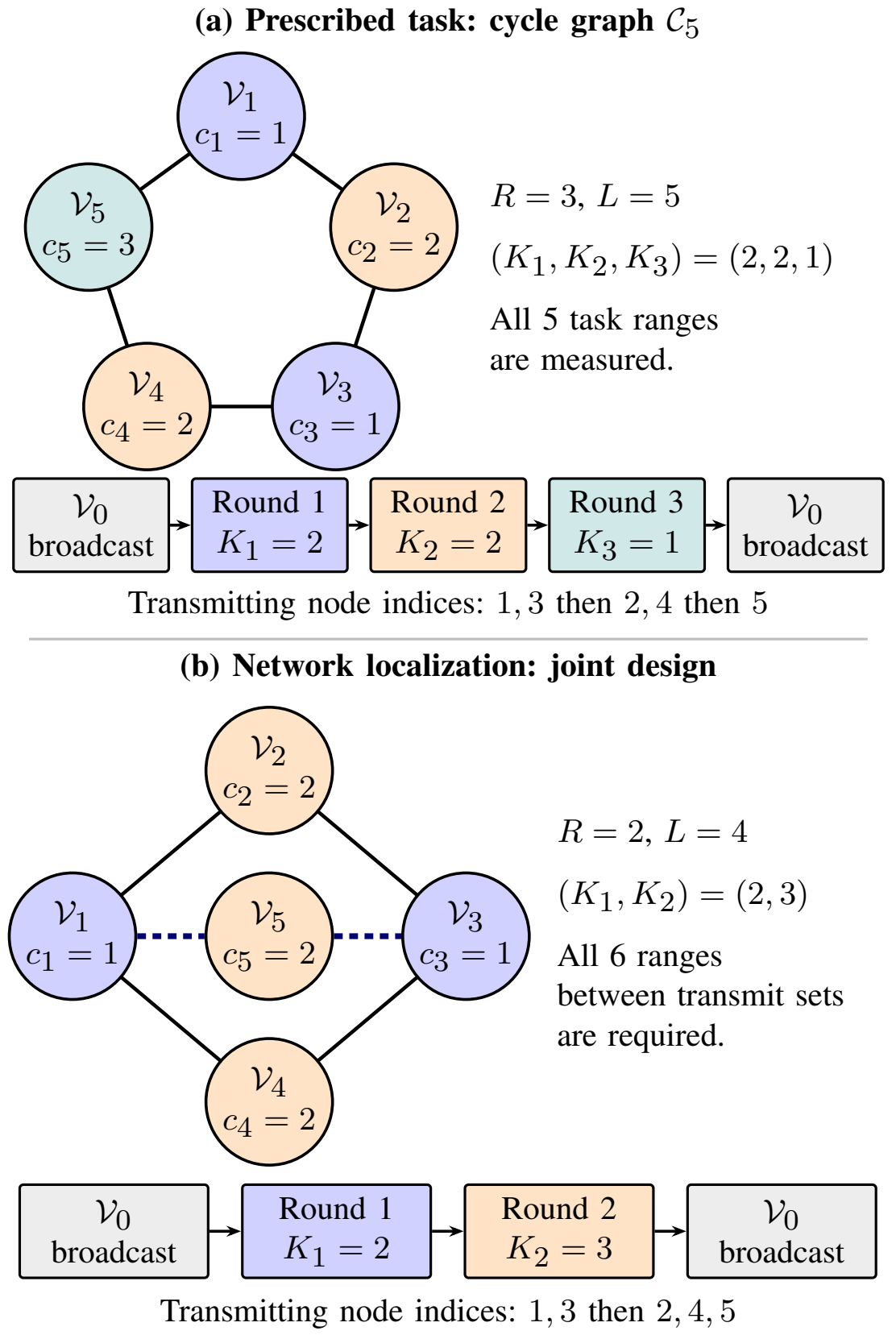
\begin{figure}[t]
\centering
\scalebox{0.85}{%
\begin{tikzpicture}[x=1cm,y=1cm,
 every node/.style={font=\fontsize{9}{11}\selectfont},
 v/.style={circle,draw,line width=.8pt,minimum size=10.2mm,
   inner sep=1pt,align=center},
 c1/.style={v,fill=blue!18},c2/.style={v,fill=orange!22},
 c3/.style={v,fill=teal!18},
 added edge/.style={draw=blue!45!black,line width=1.6pt,
   dash pattern=on 2.4pt off 1.8pt,dash phase=0pt,line cap=butt},
 slot/.style={draw,line width=.7pt,rounded corners=1pt,
   minimum width=1.82cm,minimum height=.9cm,align=center,inner sep=2pt},
 aslot/.style={slot,minimum width=1.5cm,inner sep=1pt}]
\path[use as bounding box](0,0)rectangle(8.5,13.03);
\node[font=\fontsize{9}{11}\selectfont\bfseries]at(4.25,12.73)
 {(a) Prescribed task: cycle graph $\mathcal C_5$};
\begin{scope}[shift={(2.25,10.28)}]
 \foreach \n in {1,...,5}{
  \coordinate(a\n)at({90-72*(\n-1)}:1.55);}
 \draw[line width=1pt](a1)--(a2)--(a3)--(a4)--(a5)--(a1);
 \foreach \n/\colorid in {1/1,2/2,3/1,4/2,5/3}{
  \node[c\colorid]at(a\n){$\mathcal V_{\n}$\\$c_{\n}=\colorid$};}
\end{scope}
\node[align=left,text width=3.8cm,inner sep=0pt]at(6.55,10.28)
 {$R=3$, $L=5$\\[6pt]
  $(K_1,K_2,K_3)=(2,2,1)$\\[6pt]
  All 5 task ranges\\are measured.};
\node[aslot,fill=black!7](as0)at(.8,7.88){$\mathcal V_0$\\broadcast};
\node[aslot,fill=blue!18](as1)at(2.525,7.88){Round 1\\$K_1=2$};
\node[aslot,fill=orange!22](as2)at(4.25,7.88){Round 2\\$K_2=2$};
\node[aslot,fill=teal!18](as3)at(5.975,7.88){Round 3\\$K_3=1$};
\node[aslot,fill=black!7](as4)at(7.7,7.88){$\mathcal V_0$\\broadcast};
\draw[-{Stealth[length=1.4mm]},line width=.8pt](as0)--(as1);
\draw[-{Stealth[length=1.4mm]},line width=.8pt](as1)--(as2);
\draw[-{Stealth[length=1.4mm]},line width=.8pt](as2)--(as3);
\draw[-{Stealth[length=1.4mm]},line width=.8pt](as3)--(as4);
\node at(4.25,7.10){Transmitting node indices: $1,3$ then $2,4$ then $5$};
\draw[black!25,line width=.7pt](.2,6.78)--(8.3,6.78);
\node[font=\fontsize{9}{11}\selectfont\bfseries]at(4.25,6.50)
 {(b) Network localization: joint design};
\begin{scope}[shift={(2.52,3.91)}]
 \coordinate(b1)at(-1.9,0);\coordinate(b3)at(1.9,0);
 \coordinate(b2)at(0,1.6);\coordinate(b4)at(0,-1.6);
 \coordinate(b5)at(0,0);
 \draw[line width=1pt](b1)--(b2)--(b3)--(b4)--(b1);
 \foreach \n/\colorid in {1/1,2/2,3/1,4/2,5/2}{
  \node[c\colorid](theta\n)at(b\n){$\mathcal V_{\n}$\\$c_{\n}=\colorid$};}
 \draw[added edge](theta1.east)--(theta5.west);
 \draw[added edge](theta5.east)--(theta3.west);
\end{scope}
\node[align=left,text width=3.2cm,inner sep=0pt]at(6.9,3.91)
 {$R=2$, $L=4$\\[6pt]
  $(K_1,K_2)=(2,3)$\\[6pt]
  All 6 ranges\\between transmit sets\\are required.};
\node[slot,fill=black!7](s0)at(1.02,1.06){$\mathcal V_0$\\broadcast};
\node[slot,fill=blue!18](s1)at(3.17,1.06){Round 1\\$K_1=2$};
\node[slot,fill=orange!22](s2)at(5.32,1.06){Round 2\\$K_2=3$};
\node[slot,fill=black!7](s3)at(7.47,1.06){$\mathcal V_0$\\broadcast};
\draw[-{Stealth[length=1.4mm]},line width=.8pt](s0)--(s1);
\draw[-{Stealth[length=1.4mm]},line width=.8pt](s1)--(s2);
\draw[-{Stealth[length=1.4mm]},line width=.8pt](s2)--(s3);
\node at(4.25,.26){Transmitting node indices: $1,3$ then $2,4,5$};
\end{tikzpicture}
}
\caption{Prescribed cycle graph and joint design for $N=6$ and
$K_{\max}=3$, requiring 3 and 2 intermediate rounds, respectively.
Dashed edges add a path through $\mathcal V_5$ to the 4-node cycle graph.}
\label{fig:construction}
\end{figure}

Theorem~\ref{thm:joint} shows that two constraints determine the joint
minimum round count. In~\eqref{eq:joint-optimum}, the first term
accounts for the two transmit--listen rounds needed for reciprocal
coverage of a task edge. The second accounts for at least one transmission
from each of the $M$ non-reference nodes with at most $K_{\max}$
transmitters per round. Therefore, increasing the receiver capacity can
reduce the minimum round count until the 2-round limit is reached.

\begin{remark}
For $N=6$ and $K_{\max}=3$, both constructions in
Fig.~\ref{fig:construction} satisfy the localization requirement. The
prescribed $\mathcal C_5$ in panel (a) requires 3 transmit groups and
3 intermediate rounds. The joint design in panel (b) instead selects
$\mathcal K_{2,3}$, whose task edges all join 2 groups of sizes 2 and 3.
Each group transmits in one round while the other listens, so two
intermediate rounds suffice. In this context, selecting different required distances can
save one intermediate round while preserving the geometric constraints
for localization.
\end{remark}

\subsection[Minimum Number of Required Distances]{Minimum Number of Required Distances}\label{sec:minimum-distances}
We further determine how few distances are required to achieve the joint minimum round count.
The following proposition
gives the minimum edge count of the required graph $\mathcal F$,
including all reference edges.
\begin{proposition}[Sparse required graphs at the minimum round count]
\label{prop:edges}
Under Theorem~\ref{thm:joint}, consider generically globally rigid
required graphs $\mathcal F=\mathcal K_1\vee\mathcal H$ for which a feasible
schedule with $R^\star_{\rm joint}(N,K_{\max})$ intermediate rounds exists. Their minimum edge count is
\begin{equation}
 E_{\min}(N,K_{\max})=\begin{cases}
  2N-1,&M\text{ odd and }R^\star_{\rm joint}=2,\\
  2N-2,&\text{otherwise}.
 \end{cases}
 \label{eq:sparse-graph-edges}
\end{equation}
\end{proposition}
\begin{IEEEproof}
See Appendix~\ref{app:edges}.
\end{IEEEproof}
\begin{remark}
Proposition~\ref{prop:edges} shows that the minimum number of required
distances $E_{\min}(N,K_{\max})$ grows linearly with $N$ at the minimum
round count.
For example, with $N=100$ and $K_{\max}=4$, the minimum required graph
contains 198 edges, compared with 4950 for complete ranging. In this case, the joint design uses only 4\% of these distances as required edges,
while preserving the geometric constraints for localization at the
minimum round count.
With two rounds, task edges join nodes in different transmit
sets. For odd $M$, a 2-vertex-connected task graph with only $M$ edges
would be the odd cycle graph $\mathcal C_M$, which cannot satisfy this
assignment. Thereby, the minimum edge count increases by one.
\end{remark}

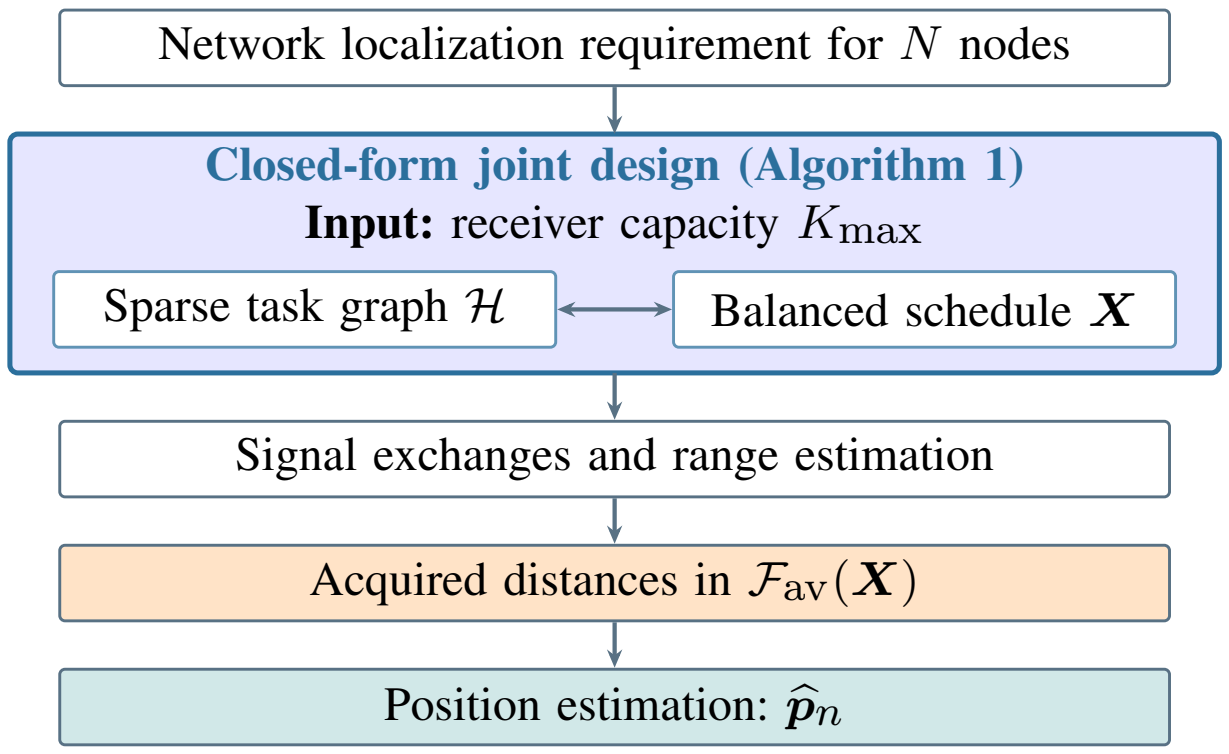
\begin{figure}[t]
\centering
\begin{tikzpicture}[x=1cm,y=1cm,
 every node/.style={font=\fontsize{9}{11}\selectfont,align=center},
 joint/.style={draw=flowblue,line width=.95pt,rounded corners=1.5pt,
               fill=blue!10},
 stage/.style={draw=flowline,line width=.55pt,rounded corners=1pt,
               fill=white,inner xsep=2pt,inner ysep=1.5pt,
               text width=7.2cm,minimum height=.50cm},
 connector/.style={draw=flowline,line width=.80pt},
 arr/.style={connector,-{Stealth[length=1.2mm,width=1.0mm]}}]
\definecolor{flowblue}{HTML}{2B6F9C}
\definecolor{flowline}{HTML}{587184}
\path[use as bounding box](0,0)rectangle(8.6,4.90);

\node[stage] (task)at(4.3,4.61)
 {Network localization requirement for $N$ nodes};

\draw[joint] (.30,2.46)rectangle(8.30,4.04);
\draw[arr](task.south)--(4.3,4.04);
\node[text=flowblue!100,font=\fontsize{9}{11}\selectfont\bfseries]
 at(4.3,3.81){Closed-form joint design (Algorithm~\ref{alg:joint})};
\node at(4.3,3.43){\textbf{Input:} receiver capacity $K_{\max}$};
\node[stage,draw=flowblue!75,text width=3.18cm]
 (graph)at(2.25,2.88){Sparse task graph $\mathcal H$};
\node[stage,draw=flowblue!75,text width=3.18cm]
 (schedule)at(6.35,2.88){Balanced schedule $\bm X$};
\draw[connector,{Stealth[length=1.2mm,width=1.0mm]}-{Stealth[length=1.2mm,width=1.0mm]}]
 (graph.east)--(schedule.west);

\node[stage] (ranging)at(4.3,1.89)
 {Signal exchanges and range estimation};
\draw[arr](4.3,2.46)--(ranging.north);
\node[stage,fill=orange!22] (available)at(4.3,1.07)
 {Acquired distances in $\mathcal F_{\rm av}(\bm X)$};
\draw[arr](ranging.south)--(available.north);
\node[stage,fill=teal!18] (estimate)at(4.3,.25)
 {Position estimation: $\widehat{\bm p}_n$};
\draw[arr](available.south)--(estimate.north);
\end{tikzpicture}
\caption{An illustration of the task-oriented S-SM-NR framework.}
\label{fig:flow}
\end{figure}

\section{Task-Oriented Ranging Schedule Design}\label{sec:algorithms}
In this section, we develop a closed-form joint design that achieves
the minimum intermediate-round count in Theorem~\ref{thm:joint}.
In Section~\ref{sec:scheduling-overview}, we introduce the design
objective and the ranging scheme. In Section~\ref{sec:closed-form-design},
we construct the task graph and assign each non-reference node a transmit round.
In Section~\ref{sec:schedule-formation}, we use these assignments to form the
schedule and determine the acquired distances.
Section~\ref{sec:complexity} analyzes the time and space
complexity of the closed-form design.

\subsection{Overview of the Scheduling Framework}\label{sec:scheduling-overview}
In Fig.~\ref{fig:flow}, we outline the S-SM-NR framework from the
localization requirement to position estimation. Within this framework,
Algorithm~\ref{alg:joint} jointly selects a sparse task graph
$\mathcal H$ and its transmit--listen schedule $\bm X$ to shorten the
ranging cycle. The key idea is to retain enough distances for localization
while choosing task edges that can be measured within the minimum
round count. We use a cycle graph as the starting point because it
remains connected after any one node is removed. Together with the
reference edges, this connectivity property inherently ensures global rigidity by
Lemma~\ref{lem:cone}.
The construction is feasible whenever the physical graph
contains every edge of the constructed task graph $\mathcal H$.

In the designed protocol, each non-reference node transmits in one round and listens in the other
rounds. Therefore, we assign the endpoints of every task edge to
different rounds, so each endpoint can receive from the other. At the same time, we balance the number of transmitters across rounds to keep
the round loads within the receiver capacity $K_{\max}$. The task graph and round
assignments are optimized together to meet these requirements within
$R^\star_{\rm joint}(N,K_{\max})$ rounds in Theorem~\ref{thm:joint}.
The required graph also achieves the minimum edge count
established in Proposition~\ref{prop:edges}.
The resulting graph and schedule determine the
signal exchanges, and all acquired distances in
$\mathcal F_{\rm av}(\bm X)$ are used for position estimation.

\subsection[Task Graph Construction and Round Assignment]{Task Graph Construction and Round Assignment}\label{sec:closed-form-design}

\begin{algorithm}[t]
	\caption{Closed-form graph and schedule design}\label{alg:joint}
	\small
\begin{algorithmic}[1]
		\Require Receiver capacity $K_{\max}$
		\State \textbf{Stage 1: Construct task graph and assign rounds}
		\State $M\gets N-1$ and
		$R\gets\max\{2,\lceil M/K_{\max}\rceil\}$.
		\If{$R=2$ and $M$ is odd}
		\State $\mathcal H\gets\mathcal C_{M-1}$.
		\State $c_n\gets1+((n-1)\bmod2)$ for $1\le n<M$.
		\State Add vertex $\mathcal V_M$ and set
		$\mathcal E_{\mathcal H}\gets\mathcal E_{\mathcal H}\cup
		\{\mathcal E_{M,1},\mathcal E_{M,3}\}$.
		\State $c_M\gets2$.
		\Else
		\State $\mathcal H\gets\mathcal C_M$.
		\State
		$c_n\gets1+((n-1)\bmod R)$ for $1\le n\le M$.
		\If{$M\bmod R=1$}
		\State $(c_{M-1},c_M)\gets(c_M,c_{M-1})$.
		\EndIf
		\EndIf
		\State \textbf{Stage 2: Form the schedule}
		\State $\mathcal A_n\gets\{c_n\}$ for $1\le n\le M$.
		\State $x_{n,r}\gets\mathbb I\{c_n=r\}$ for $1\le n\le M$, $1\le r\le R$.
		\State Place the reference broadcasts before round 1 and after round $R$ for the reference node $\mathcal V_0$.
		\Ensure Task graph $\mathcal H$ and transmit--listen schedule $\bm X$.
	\end{algorithmic}
\end{algorithm}

We use $R=R^\star_{\rm joint}(N,K_{\max})$
from~\eqref{eq:joint-optimum}. In Stage 1 of Algorithm~\ref{alg:joint},
we select the task graph and assign each non-reference node a
transmit-round index $c_n\in\{1,\cdots,R\}$. We use a cycle graph when
$R\ge3$ or $M$ is even. When $R=2$ and $M$ is odd, one should modify the task
graph to allow a two-round schedule.

\subsubsection[Cycle Graph and Round Assignment in the General Case]{Cycle Graph and Round Assignment in the General Case}\label{sec:general-construction}
First, we form the cycle graph $\mathcal C_M$ with $M$ task edges
by connecting the non-reference nodes in order and adding the closing
edge between $\mathcal V_M$ and $\mathcal V_1$. The closing edge is
essential for the localization condition. A path would split if an
interior node were removed, whereas removing any one node from the
cycle leaves a connected path through the others. By
Lemma~\ref{lem:cone}, combining these task edges with the reference
edges gives a required graph satisfying the global-rigidity condition.

Then, we assign the round indices
$1,\cdots,R$ repeatedly along the node order of $\mathcal C_M$.
In this way, the successive nodes from $\mathcal V_1$ to $\mathcal V_M$ transmit
in different rounds, which allows each adjacent node pair to exchange signals in both
directions. This repeated assignment also distributes the nodes evenly
across the rounds, with round loads that differ by at most one. Each round
contains at most $\lceil M/R\rceil$ transmitters, which is no larger
than $K_{\max}$ for the chosen $R$.

Finally, we check whether the endpoints of the closing edge
transmit in different rounds. If $M\bmod R=1$,
node $\mathcal V_M$ and node $\mathcal V_1$ both transmit in round 1 and
cannot listen to each other. We exchange the assignments of
$\mathcal V_{M-1}$ and $\mathcal V_M$. The final three indices become
$R-1$, $1$, and $R$, so the endpoints of the closing edge
and of each other affected task edge transmit in different rounds. The exchange leaves the round loads unchanged, so the
receiver-capacity limit is preserved and every task edge has reciprocal
coverage.

\subsubsection[Graph Adjustment in the Special Two-Round Case]{Graph Adjustment in the Special Two-Round Case}\label{sec:two-round-construction}
When $M$ is odd and $R=2$, alternating rounds 1 and 2 around
$\mathcal C_M$ always assigns the endpoints of the closing edge to
the same round. To achieve the minimum round count,
the task graph should be modified. First, we form the even cycle graph $\mathcal C_{M-1}$
and alternate the two round indices along it. Then, we connect the
remaining node $\mathcal V_M$ to $\mathcal V_1$ and $\mathcal V_3$,
both of which transmit in round 1. Finally, we assign $\mathcal V_M$
to round 2, which allows us to measure both added distances
without adding another round.

The two connections preserve the localization condition when the
remaining node is added. With only one connection, removing its neighbor
would disconnect the added node from the task graph. With two distinct
neighbors on the cycle, the task graph remains connected after any
one node is removed. The round loads are $(M-1)/2$ and $(M+1)/2$, both of them are at most $K_{\max}$.

\begingroup

\begin{remark}
To illustrate the special two-round construction, we consider $M=5$ and $K_{\max}=3$,
for which the minimum round count is $R=2$. In Fig.~\ref{fig:construction}(b), nodes
$\mathcal V_1,\cdots,\mathcal V_4$ form an even cycle graph.
The dashed edges connect the remaining node $\mathcal V_5$ to
$\mathcal V_1$ and $\mathcal V_3$. These two neighbors transmit in
round 1, while $\mathcal V_5$ transmits with $\mathcal V_2$ and
$\mathcal V_4$ in round 2. In this way, node $\mathcal V_5$ receives
the signals of both neighbors in the first round and sends its signal
to them in the second. The round loads are 2 and 3, both within the
receiver capacity.
\end{remark}
\endgroup

\subsection[Schedule Formation and Available Distances]{Schedule Formation and Available Distances}\label{sec:schedule-formation}
In Stage 2 of Algorithm~\ref{alg:joint}, we use the round indices
from either construction to form the complete schedule matrix $\bm X$
by setting $x_{n,r}=\mathbb I\{c_n=r\}$ for every node and round.
Each node transmits once and listens in the other rounds. Substituting
this assignment into~\eqref{eq:gamma}, we have
\begingroup

\begin{equation}
 \Gamma_{n_1\to n_2}(\bm X)
 =\Gamma_{n_2\to n_1}(\bm X)
 =\mathbb I\{c_{n_1}\ne c_{n_2}\}.
 \label{eq:round-index-coverage}
\end{equation}
\endgroup
Thus, the endpoints of every task edge have one observation
opportunity in each direction. Then, we place the reference broadcasts before
and after these $R$ rounds to complete the ranging schedule.

In addition, the same schedule acquires distances beyond the selected task
edges. Under complete physical connectivity, the available non-reference
edges join exactly the nodes assigned to different rounds. By counting
these edges and the $M$ reference edges, with
$\sum_{r=1}^{R}K_r=M$, we have
\begin{equation}
 |\mathcal E_{\mathcal F_{\rm av}(\bm X)}|
       =M+\frac12\left(M^2-\sum_{r=1}^{R}K_r^2\right).
 \label{eq:density}
\end{equation}
To see how the round loads affect the number of available
distances, we rewrite~\eqref{eq:density} as
\begingroup

\begin{equation}
 \begin{aligned}
 |\mathcal E_{\mathcal F_{\rm av}(\bm X)}|
 =M+\frac{M^2}{2}\left(1-\frac{1}{R}\right)
 -\frac{1}{2}\sum_{r=1}^{R}
          \left(K_r-\frac{M}{R}\right)^2.
 \end{aligned}
 \label{eq:load-balance-distances}
\end{equation}
\endgroup

	For fixed $M$ and $R$, the balanced round loads in
	Algorithm~\ref{alg:joint} minimize the sum of squared deviations from
	the average load $M/R$, which maximizes the number of available
	distances. For example, with $N=12$ and $K_{\max}=5$, the minimum
	intermediate-round count is $R=3$. Both load assignments $(4,4,3)$
	and $(5,5,1)$ can cover the cycle task graph $\mathcal C_{11}$
	within the receiver-capacity limit, with each non-reference node
	transmitting once. However, the balanced assignment provides
	51 available distances, whereas the unbalanced assignment provides
	46. In this case, load balancing supplies 5 additional distances without
	increasing the round count. These additional distances can be used
	together with those required by the task graph $\mathcal H$ for the localization of the network nodes.

\subsection{Time and Space Complexity Analysis}\label{sec:complexity}
In terms of space complexity, the task graph $\mathcal H$ has $M$
or $M+1$ edges, and each non-reference node has one transmit-round
index $c_n$. The edge list and round assignments require
${\cal O}(N)$ storage. Stage 2 forms the complete
$M\times R$ schedule matrix $\bm X$, which requires ${\cal O}(MR)$ storage.
Overall, the total space complexity of
Algorithm~\ref{alg:joint}, including its outputs, is
${\cal O}(N+MR)={\cal O}(MR)$.

In terms of time complexity, Algorithm~\ref{alg:joint} computes $c_n$
and adds a constant number of task edges per node. The closing-edge
adjustment or the insertion of the remaining node takes constant time.
Therefore, Stage 1 requires ${\cal O}(N)$ time. In Stage 2,
forming the $M\times R$ schedule matrix requires ${\cal O}(MR)$ time. Overall, the total time complexity of
Algorithm~\ref{alg:joint} is ${\cal O}(N+MR)={\cal O}(MR)$.
For fixed $K_{\max}$, $R$ grows linearly with $N$, so the total time
and space complexities are ${\cal O}(N^2)$.

\section{Numerical Results}\label{sec:numerical}
In this section, numerical simulations are carried out to demonstrate the efficiency and accuracy of the proposed S-SM-NR framework.

\subsection{Simulation Setup}\label{sec:setup}
To compare different ranging schemes under the same network conditions,
we use a common timestamp-error model and position estimator. The position accuracy is measured by the root-mean-square error (RMSE).

\subsubsection{Network Conditions}
The initial positions of nodes $\mathcal V_3,\cdots,\mathcal V_M$ are uniformly distributed in a $300\times300$~m$^2$ area.
Unless otherwise stated, we use $N=100$ nodes, the receiver capacity
$K_{\max}=4$, and the maximum node speed $v_{\max}=3$~m/s.
The node speeds are independent and uniformly
distributed over $[0,v_{\max}]$, and the directions are independent and
uniformly distributed over $[0,2\pi)$. Each node maintains a constant speed and direction during the ranging cycle.
All schemes start their ranging cycle simultaneously, and the time interval between two adjacent rounds is 50~ms.

For the network topology, all ranging schemes use the same physical connectivity graph $\mathcal G_{\rm phy}$, which contains all reference edges and a graph in which each non-reference node has 4 non-reference neighbors. The graph contains $3(N-1)$ distances as $N$ varies. For the
default setup, we use cyclic node indices and connect each non-reference
node to the nodes at offsets $\pm1$ and $\pm38$.\footnote{For
$N=20$, 40, 60, 80, 120, 140, and 160, the offset pairs are $(1,3)$,
$(1,14)$, $(1,19)$, $(1,5)$, $(1,13)$, $(1,51)$, and $(1,17)$,
respectively. Each pair specifies offsets in both directions on the
non-reference cycle graph.} The offset-1 edges form a cycle
graph, which satisfies the connectivity condition in
Lemma~\ref{lem:cone}.
The physical graph contains the task graph constructed by
Algorithm~\ref{alg:joint}, so the joint minimum round count can be
realized in this sparse network. Throughout the receiver-capacity sweep from 1 to 16, the schedule of
each scheme provides signal exchanges in both directions for every physically connected node pair. In this way, each scheme has the ability to acquire all distances in $\mathcal G_{\rm phy}$, giving
$\mathcal F_{\rm av}(\bm X)=\mathcal G_{\rm phy}$.

\subsubsection{Compared Schemes}
 The three ranging schemes to be compared are described as follows.

\begin{enumerate}
\item \textbf{SM-NR~\cite{zhang2022smnr}:} In the SM-NR scheme, all nodes
transmit sequentially, following its complete-ranging design. The schedule
requires $N+1$ occupied rounds in each ranging cycle.
\item \textbf{CC-SM-NR~\cite{ccsmnr2026}:} In the CC-SM-NR scheme, all
nodes use distinct transmit--listen codewords for concurrent transmissions. The common codeword weight $w$
specifies the number of transmit rounds per non-reference node. Within
this constant-weight codeword family, the scheme selects the smallest $R$ for which a weight $w$
satisfies $\binom{R}{w}\ge M$ and $Mw\le RK_{\max}$.
\item \textbf{Proposed S-SM-NR:} Algorithm~\ref{alg:joint} is adopted to construct the
cycle task graph as well as the balanced transmit--listen schedule for network localization.
\end{enumerate}

\subsubsection[Timestamp Modeling and Range Estimation]{Timestamp Modeling and Range Estimation}
To generate range observations under asynchronous clocks, we model the
timestamp error of each physical event. For an event $\alpha$ at true
time $t_\alpha$, node $\mathcal V_n$ with clock skew $\rv{e}_n$ records
\begin{equation}
  \rv{\gamma}_\alpha^{(n)}
  = (1+\rv{e}_n)t_\alpha+b_n+\rv{\eta}_\alpha^{(n)},
  \label{eq:timestamp-model}
\end{equation}
where $\rv{e}_n$ describes the clock-frequency deviation, $b_n$ is the
clock offset, and $\rv{\eta}_\alpha^{(n)}$ is the event timestamp error.
The TX and RX timestamp errors are zero-mean Gaussian
with standard deviations $\sigma$ and $\sigma\sqrt{K_r}$,
respectively~\cite{ccsmnr2026}. The default value of the timestamp-error standard deviation is set to
$\sigma=0.05$~ns, while the clock-frequency deviations $\{\rv{e}_n\}_{n=0}^M$ are uniformly distributed within $[-40,40]$~ppm.

From these timestamps, the range estimate
$\widehat d_{n_1,n_2}$ can be obtained by adopting the clock-frequency compensation and reciprocal
ranging method in~\cite[eqs.~(8), (18), and~(19)]{ccsmnr2026}. Then, the resulting distances are used for position estimation.

\begin{figure}[t]\centering
	\includegraphics[width=\simroundwidth]{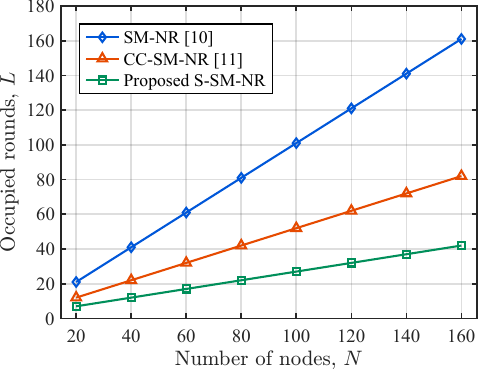}
	\caption{Occupied rounds per ranging cycle versus network size.}
	\label{fig:roundN}\end{figure}

\subsubsection[Position Estimation and Accuracy Evaluation]{Position Estimation and Accuracy Evaluation}
We estimate the node positions by nonlinear least squares.
To compare the estimates in a common coordinate frame, we fix
the positions of $\mathcal V_0$, $\mathcal V_1$, and $\mathcal V_2$ in
the estimator at their known initial coordinates
$(0,0)$~m, $(300,0)$~m, and $(0,300)$~m, respectively.
Let ${\widehat{\bm p}}_n \in\mathbb R^2$ denote the estimated position of node
$\mathcal V_n$, with ${\widehat{\bm p}}_n=\bm p_n$ for $n=0,1,2$.
The estimation problem is
\begingroup
\begin{equation}
 \min\limits_{{\widehat{\bm p}}_3,\cdots,{\widehat{\bm p}}_M}
 \sum_{\mathcal E_{n_1,n_2}\in\mathcal E_{\mathcal F_{\rm av}(\bm X)}}
 \left(\| {\widehat{\bm p}}_{n_1}-{\widehat{\bm p}}_{n_2}\|-\widehat d_{n_1,n_2}\right)^2.
 \label{eq:position-objective}
\end{equation}
\endgroup
For all three ranging schemes, the least-square estimation problem in ~\eqref{eq:position-objective} can be iteratively solved with the
damped Gauss--Newton/Levenberg--Marquardt solver~\cite{more1978levenberg}.
Then, the position RMSE is computed
over the remaining $N-3$ nodes against their true positions at the common
start of ranging as
\begin{equation}
 \mathrm{RMSE}
 =\sqrt{\mathbb E\!\left[
 \frac{1}{N-3}\sum_{n=3}^{M}
 \left\|\widehat{\bm p}_n-\bm p_n\right\|^2
 \right]},
 \label{eq:position-rmse}
\end{equation}
where $\mathbb E[\cdot]$ averages over the simulation realizations.

\begin{figure}[t]\centering
	\includegraphics[width=\simroundwidth]{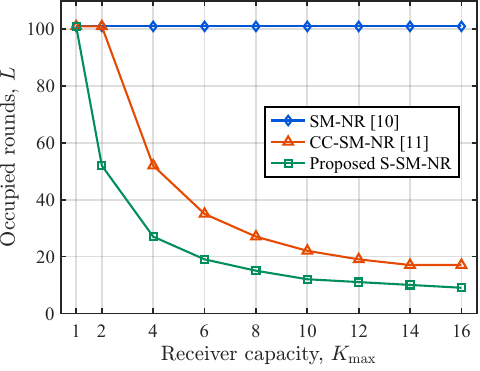}
	\caption{Occupied rounds per ranging cycle versus receiver capacity.}
	\label{fig:roundK}\end{figure}

\subsection{Round Count of Different Ranging Schemes}\label{sec:ranging-duration}
In Fig.~\ref{fig:roundN} and Fig.~\ref{fig:roundK}, we plot the occupied-round
count $L$ as the network size and receiver capacity vary. The S-SM-NR
scheme follows the joint minimum-round result in Theorem~\ref{thm:joint}, i.e.,
\begin{equation}
 L=\max\left\{2,\left\lceil\frac{N-1}{K_{\max}}\right\rceil\right\}+2.
 \label{eq:cycle-round-count}
\end{equation}
At the default setting, the S-SM-NR scheme occupies 27 rounds, compared with 101 for
the SM-NR scheme and 52 for the CC-SM-NR scheme. The S-SM-NR scheme assigns one transmission to each
non-reference node and achieves the capacity lower bound of 25
intermediate rounds. The CC-SM-NR scheme assigns two transmissions per non-reference
node and requires 50 intermediate rounds under the same receiver capacity.

One can find that both concurrent schemes make nearly full use
of the receiver capacity. With fewer transmissions per non-reference node, S-SM-NR completes
the ranging task in fewer rounds than CC-SM-NR.
The savings in occupied rounds increase with the network size.
Every 20 additional nodes require 5 additional rounds for S-SM-NR,
compared with 20 for SM-NR and 10 for CC-SM-NR. By allowing nodes
without a required mutual range to share a transmit round, the S-SM-NR scheme
limits the growth of the ranging cycle while retaining the distances needed for localization.

Besides, in Fig.~\ref{fig:roundK}, increasing $K_{\max}$ from 1 to 2 reduces
the S-SM-NR round count from 101 to 52, whereas the SM-NR and CC-SM-NR
schemes remain at 101 rounds. The sparse task graph permits concurrent
transmission with one transmission per non-reference node, allowing
S-SM-NR to exploit even this small capacity increase. In contrast, the complete-ranging
codeword constraints prevent CC-SM-NR from shortening its schedule.

\subsection{Position Accuracy and Measurement Latency}\label{sec:timestamp-effects}
\begin{figure}[t]\centering
\includegraphics[width=\columnwidth]{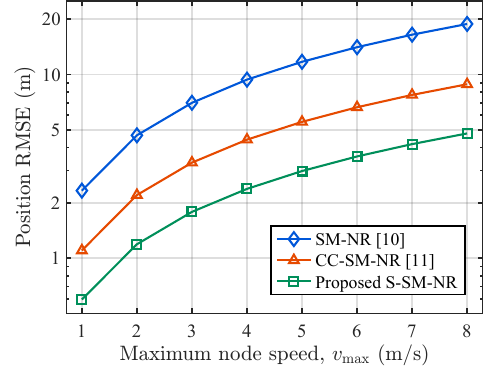}
\caption{Position RMSE versus maximum node speed.}
\label{fig:accuracy}\end{figure}

\begin{figure}[t]\centering
	\includegraphics[width=\columnwidth]{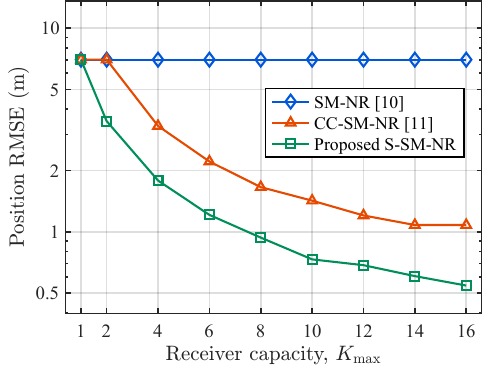}
	\caption{Position RMSE versus receiver capacity.}
	\label{fig:accuracyK}
\end{figure}

In Fig.~\ref{fig:accuracy}, we compare the position RMSE as the maximum
node speed $v_{\max}$ varies from 1~m/s to 8~m/s. The S-SM-NR
scheme gives the lowest RMSE throughout this range, with an absolute
accuracy advantage that grows with node speed. The RMSE increases
approximately in proportion to the node speed for all three schemes. At $v_{\max}=8$~m/s, the RMSEs of the SM-NR,
CC-SM-NR, and S-SM-NR schemes are approximately $18.79$~m, $8.83$~m,
and $4.77$~m, respectively. Thus, shortening the ranging cycle becomes
more effective as the inter-node distances change faster.

The different rates of RMSE growth reflect the effect of the
ranging-cycle duration on position accuracy. For a moving node pair,
the distance variation depends on the relative velocity along the
line connecting the nodes and the elapsed measurement time.
A longer cycle allows larger distance changes before all
measurements are acquired. Therefore, as node speed increases, the distances
collected at different stages of the cycle provide less
consistent geometric constraints for localization. This explains
the faster RMSE growth of the longer-cycle baselines. The S-SM-NR scheme
acquires the required distances within fewer rounds, which limits these changes and the resulting position error.

\subsection{Effect of Receiver Capacity}\label{sec:accuracyK}

In Fig.~\ref{fig:accuracyK}, we compare the position RMSE as $K_{\max}$
varies from 1 to 16. At $K_{\max}=1$, all three schemes use the same
sequential schedule and give the same RMSE. The RMSE of the SM-NR
scheme remains unchanged because its schedule is fixed. Increasing
$K_{\max}$ from 4 to 16 reduces the S-SM-NR ranging cycle from 27 rounds
to 9 rounds and its RMSE from approximately $1.79$~m to $0.54$~m.
The shorter measurement interval improves position accuracy, consistent
with the round-count reduction in Fig.~\ref{fig:roundK}.

The CC-SM-NR curve levels off between $K_{\max}=14$ and 16
because the scheme uses the same 17-round schedule.
The RMSE remains approximately $1.08$~m.
The number of distinct codewords limits the schedule length, so
additional receiver capacity does not shorten the ranging cycle.
Over the same capacity change, S-SM-NR reduces its cycle from 10 rounds
to 9 rounds and continues to improve position accuracy. Therefore, the benefit of
greater receiver capacity depends on whether the task and schedule
design can use it to reduce the round count. Since the round count is
integer-valued, capacity gains shorten the cycle at discrete thresholds.

\subsection{Effect of Network Size}\label{sec:accuracyN}

\begin{figure}[t]\centering
	\includegraphics[width=\columnwidth]{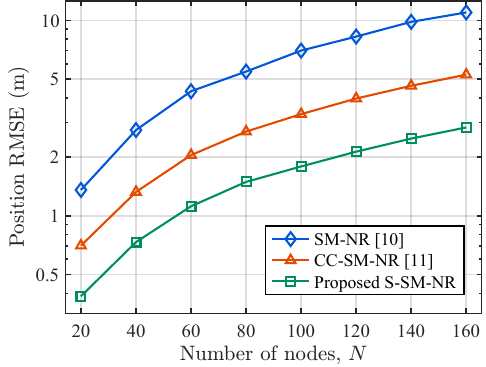}
	\caption{Position RMSE versus network size.}
	\label{fig:accuracyN}\end{figure}

In Fig.~\ref{fig:accuracyN}, we compare position accuracy as $N$ varies
from 20 to 160, with 5 available neighbors per non-reference node,
including the reference node. The S-SM-NR scheme gives the lowest RMSE
throughout this range. Its RMSE increases from approximately $0.39$~m
at $N=20$ to $2.83$~m at $N=160$, where
the SM-NR and CC-SM-NR schemes give approximately $10.98$~m and
$5.27$~m, respectively. The absolute accuracy advantage over both
baselines grows with the network size, which is consistent with the smaller
round-count growth in Fig.~\ref{fig:roundN}.

As $N$ increases, each non-reference node has the same
number of measured neighbors, so the number of measurements per
unknown node remains roughly constant. However, completing network
ranging requires more rounds, which lengthens the acquisition interval
and allows larger changes in inter-node distances during the measurement.
The resulting distances provide less consistent geometric constraints
for localization, which leads to the increase in RMSE with network
size. By sharing transmit rounds, S-SM-NR requires fewer additional
rounds as the network grows, which limits the increase in ranging-cycle
duration and position error.

\section{Conclusion}\label{sec:conclusion}
In this paper, we established S-SM-NR as a task-oriented signal-processing
framework that adapts the transmit--listen schedule to the required
ranging task. We analyzed the minimum round count and the distance requirements for network localization. Specifically, we derived minimum
round counts and lower bounds and established a connectivity condition for
network localization. Based on this analysis, we
developed a closed-form joint design of the sparse task graph and
transmit--listen schedule. By constructing task graphs from cycle graphs
and balancing round loads, the joint design achieves the derived
minimum round count.
The simulation results show that the S-SM-NR scheme reduces the occupied-round
count and improves the localization accuracy.
\label{sec:body-end}

\appendices
\section{Proof of Theorem~\ref{thm:fixed}}\label{app:fixed}
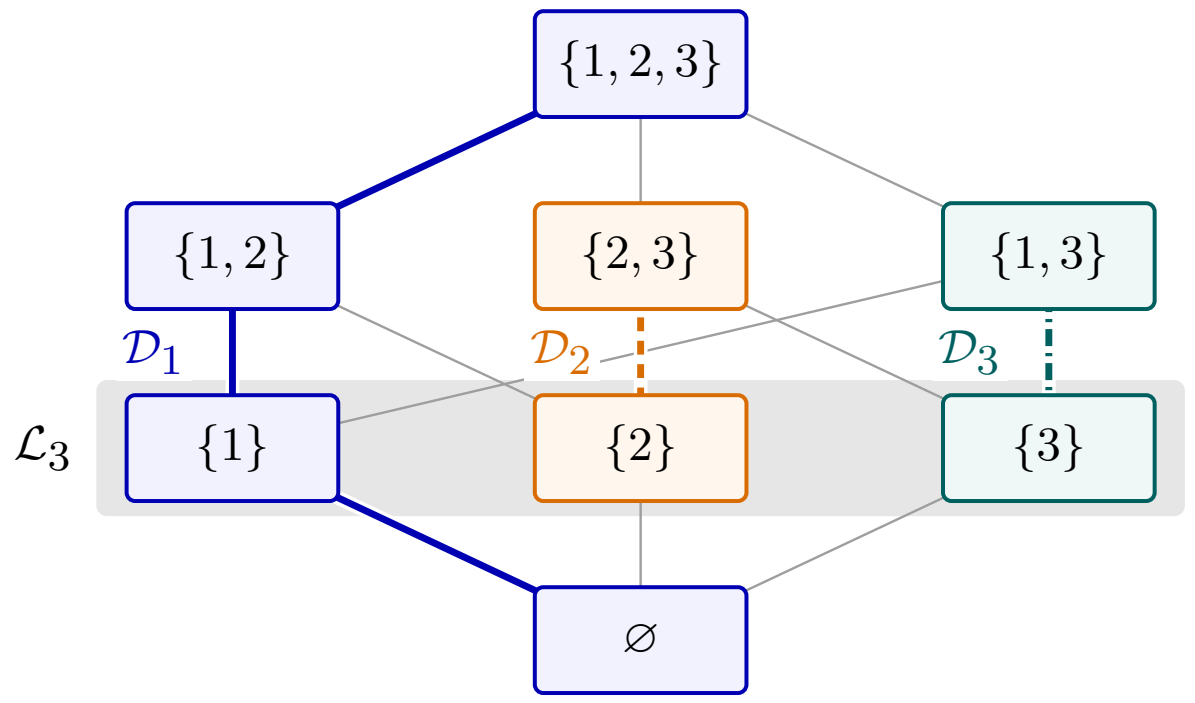
\begin{figure}[t]
\centering
\begin{tikzpicture}[x=1cm,y=1cm,
 every node/.style={font=\fontsize{9}{11}\selectfont},
 setnode/.style={rounded corners=1.6pt,line width=.75pt,
   minimum width=14mm,minimum height=7mm,inner sep=2pt,
   outer sep=0pt,text=black},
 chain/.style={line width=1.3pt,
   preaction={draw=white,line width=2.6pt,solid}},
 chainone/.style={chain,draw=blue!70!black},
 chaintwo/.style={chain,draw=orange!85!black,
   dash pattern=on 2.6pt off 1.8pt},
 chainthree/.style={chain,draw=teal!75!black,
   dash pattern=on 3.2pt off 1.3pt on .6pt off 1.3pt},
 familylabel/.style={fill=white,inner sep=1.2pt}]
\path[use as bounding box](0,0)rectangle(8.5,4.8);
\coordinate(b0)at(4.25,.50);
\coordinate(b1)at(1.55,1.77);
\coordinate(b2)at(4.25,1.77);
\coordinate(b3)at(6.95,1.77);
\coordinate(b12)at(1.55,3.04);
\coordinate(b23)at(4.25,3.04);
\coordinate(b13)at(6.95,3.04);
\coordinate(b123)at(4.25,4.31);
\fill[black!10,rounded corners=2pt](.65,1.32)rectangle(7.85,2.22);
\foreach \from/\to in
 {b0/b1,b0/b2,b0/b3,b1/b12,b1/b13,b2/b12,b2/b23,
  b3/b23,b3/b13,b12/b123,b23/b123,b13/b123}{
 \draw[black!38,line width=.45pt](\from)--(\to);}
\draw[chainone](b0)--(b1)--(b12)--(b123);
\draw[chaintwo](b2)--(b23);
\draw[chainthree](b3)--(b13);
\node[setnode,draw=blue!70!black,fill=blue!5]at(b0){$\varnothing$};
\node[setnode,draw=blue!70!black,fill=blue!5]at(b1){$\{1\}$};
\node[setnode,draw=blue!70!black,fill=blue!5]at(b12){$\{1,2\}$};
\node[setnode,draw=blue!70!black,fill=blue!5]at(b123){$\{1,2,3\}$};
\node[setnode,draw=orange!85!black,fill=orange!7]at(b2){$\{2\}$};
\node[setnode,draw=orange!85!black,fill=orange!7]at(b23){$\{2,3\}$};
\node[setnode,draw=teal!75!black,fill=teal!6]at(b3){$\{3\}$};
\node[setnode,draw=teal!75!black,fill=teal!6]at(b13){$\{1,3\}$};
\node[familylabel,text=blue!70!black]at(1.03,2.405){$\mathcal D_1$};
\node[familylabel,text=orange!85!black]at(3.73,2.405){$\mathcal D_2$};
\node[familylabel,text=teal!75!black]at(6.43,2.405){$\mathcal D_3$};
\node at(.3,1.77){$\mathcal L_3$};
\end{tikzpicture}
\caption{Transmit supports for $R=3$. Along each colored path, each set is contained in the next. The shaded supports each contain only one round. None of these supports is contained in another. At most 3 supports
can be selected with this property.}
\label{fig:boolean}
\end{figure}

First, we derive a lower bound on the minimum round count
from any feasible schedule. Then, we construct a schedule that achieves it.
Consider any feasible $R$-round schedule and the transmit
supports assigned to its nodes. The subsets of $\mathcal R$ can be partitioned into
$\binom{R}{\lfloor R/2\rfloor}$ families, with the sets in each family
arranged so that each set is contained in the next~\cite{vanlint2001combinatorics}.
We assign each node the label of the family containing its transmit
support. For any two supports in the same family, one is contained in
the other, so their nodes cannot be joined by a task edge under
Proposition~\ref{prop:reciprocal}. Therefore, the family labels give a proper coloring of
$\mathcal H$, since each task edge joins nodes with different labels.
This coloring uses no more groups than the number of families, so we obtain
\begin{equation}
 \chi(\mathcal H)\le\binom{R}{\left\lfloor\frac{R}{2}\right\rfloor}.
 \label{eq:app-chromatic-bound}
\end{equation}
Hence, $R\ge\rho(\chi(\mathcal H))$.
In Fig.~\ref{fig:boolean}, we illustrate this partition for $R=3$,
where the 8 transmit supports are arranged into 3 families
$\mathcal D_1$, $\mathcal D_2$, and $\mathcal D_3$.
Using these three families as group labels gives
$\chi(\mathcal H)\le3$ for any task graph that admits a 3-round schedule.
Along each colored path, every set is contained in the next one.
For example, $\{1\}\subset\{1,2\}$ in $\mathcal D_1$.
If two nodes use these supports, the node with support $\{1\}$ never
transmits while the other listens. In this case, these two nodes cannot form a required node pair.

To achieve the round-count lower bound, we set
$C=\chi(\mathcal H)$ and $R=\rho(C)$.
There are at least $C$ distinct supports of size $\lfloor R/2\rfloor$,
so a different support can be assigned to each group of a proper coloring
with $C$ groups.
The endpoints of every task edge have distinct supports
of equal size. Neither support can contain the other, since containment
between sets of equal size would make them identical.
Therefore, each endpoint of a task edge has at least one
transmit round in which the other endpoint listens. These rounds
provide reciprocal coverage of the task edge.
The shaded row $\mathcal L_3$ in Fig.~\ref{fig:boolean} contains the
supports $\{1\}$, $\{2\}$, and $\{3\}$ of size 1. For $C=3$, each group transmits in a different round.
Each group listens in the other two rounds, so every required node pair can exchange signals in both directions.

If $\mathcal H$ has an edge, then $C\ge2$ and $R\ge2$, so all assigned
supports are nonempty. If $\mathcal H$ has no edges, reciprocal coverage
needs no rounds. If every non-reference node is required to transmit at least once,
the minimum is one round under unlimited receiver capacity. This proves both statements.

\section{Proof of Proposition~\ref{prop:finite}}\label{app:bounds}
We derive lower bounds on the round count by counting the
transmissions and directed observations required by the task graph.
Theorem~\ref{thm:fixed} gives the unlimited-capacity lower
bound. Since all $M$ non-reference nodes transmit at least once and each
round allows at most $K_{\max}$ transmitters, we also have
$R\ge\lceil M/K_{\max}\rceil$. Taking the maximum of these two bounds
gives~\eqref{eq:bounds-basic}. Under half-duplex operation, the observations required in both
directions of each task edge take place in different rounds, which
gives~\eqref{eq:bounds-halfduplex}.

To prove~\eqref{eq:bounds-degree}, we consider a node with $D_n>0$
task neighbors. Each of the $D_n$ task neighbors needs to transmit at least
once while node $\mathcal V_n$ listens. After counting the transmissions in these listening rounds, we have
\begingroup
\begin{equation}
 D_n\le\sum_{r=1}^{R}(1-x_{n,r})K_r
    \le K_{\max}(R-w_n)\le K_{\max}(R-1).
 \label{eq:app-listening-capacity}
\end{equation}
\endgroup
The last two inequalities use $K_r\le K_{\max}$ and
$w_n\ge1$, which gives~\eqref{eq:bounds-degree}.

Next, we bound the rounds required to cover the task edges within a
node subset $\mathcal W$. To count the directed observation opportunities
available in each round, we define
\begin{equation}
 b_{K_{\max}}(\mathcal W)=
 \max\limits_{\substack{\mathcal S\subseteq\mathcal W\\|\mathcal S|\le K_{\max}}}
 |\mathcal B_{\mathcal H[\mathcal W]}(\mathcal S)|,
 \label{eq:cut-capacity}
\end{equation}
where $\mathcal B_{\mathcal H[\mathcal W]}(\mathcal S)$ contains the
task edges joining transmitters in $\mathcal S$ to listeners in
$\mathcal W\setminus\mathcal S$. In one round, each such edge provides an observation
from its transmitting endpoint to its listening endpoint. An edge with
both endpoints transmitting or both listening provides no observation
in that round. By~\eqref{eq:cut-capacity}, at most
$Rb_{K_{\max}}(\mathcal W)$ observation opportunities are available
over $R$ rounds. Each task edge needs observations in both directions,
so we require
$2|\mathcal E_{\mathcal H[\mathcal W]}|\le Rb_{K_{\max}}(\mathcal W)$.
If $\mathcal H[\mathcal W]$ has an edge, then
$b_{K_{\max}}(\mathcal W)>0$, which allows us to divide by $b_{K_{\max}}(\mathcal W)$
and round up to obtain~\eqref{eq:bounds-cut}.
This completes the proof.

\section{Proof of Lemma~\ref{lem:cone}}\label{app:cone}
For a graph with at least four vertices, generic global
rigidity in the plane is equivalent to 3-vertex-connectivity and
redundant rigidity~\cite{jackson2005}. To prove the lemma, we check
whether the graph stays connected after deleting at most two vertices
and stays generically rigid after deleting any one edge. Rigidity rules
out continuous changes of shape that preserve the edge distances.

\subsection{Necessity of Two-Vertex Connectivity}
To prove necessity, we suppose that $\mathcal F$ is generically globally
rigid and hence 3-vertex-connected. After $\mathcal V_0$ is deleted,
$\mathcal H$ remains connected. If $\mathcal V_0$ and any non-reference
vertex $\mathcal V_n$ are deleted together, $\mathcal H-\mathcal V_n$
also remains connected.
Thus, $\mathcal H$ is 2-vertex-connected.

\subsection{Rigidity Provided by the Reference Edges}
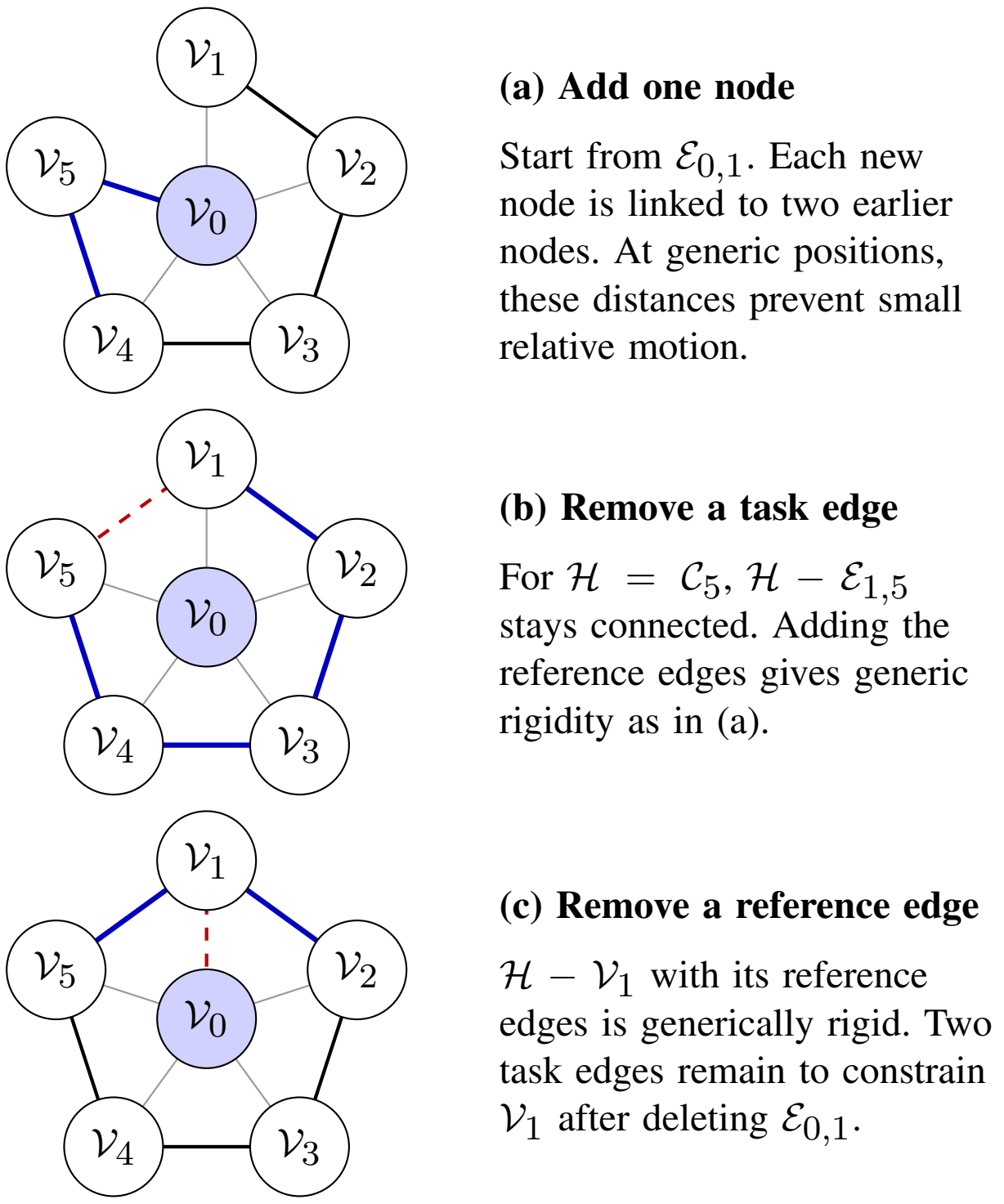
\begin{figure}[t]
\centering
\begin{tikzpicture}[x=1cm,y=1cm,
 every node/.style={font=\fontsize{9}{11}\selectfont},
 v/.style={circle,draw,fill=white,minimum size=8mm,inner sep=1pt},
 explanation/.style={align=left,text width=4.1cm}]
\path[use as bounding box](0,0)rectangle(8.5,10.1);
\foreach \panel/\height in {1/8.25,2/5,3/1.75}{
 \begin{scope}[shift={(1.8,\height)}]
  \coordinate(h\panel)at(0,0);
  \foreach \n in {1,...,5}{
   \coordinate(v\panel\n)at({90-72*(\n-1)}:1.28);}
  \ifnum\panel=1
   \draw[thick](v11)--(v12)--(v13)--(v14);
   \foreach \n in {1,...,4}{\draw[black!40](h1)--(v1\n);}
   \draw[blue!75!black,line width=1.2pt](v14)--(v15)--(h1);
  \fi
  \ifnum\panel=2
   \foreach \n in {1,...,5}{\draw[black!40](h2)--(v2\n);}
   \draw[blue!75!black,line width=1.2pt]
    (v21)--(v22)--(v23)--(v24)--(v25);
   \draw[red!75!black,dashed,thick](v25)--(v21);
  \fi
  \ifnum\panel=3
   \draw[thick](v32)--(v33)--(v34)--(v35);
   \foreach \n in {2,...,5}{\draw[black!40](h3)--(v3\n);}
   \draw[red!75!black,dashed,thick](h3)--(v31);
   \draw[blue!75!black,line width=1.2pt](v32)--(v31)--(v35);
  \fi
  \node[v,fill=blue!18]at(h\panel){$\mathcal V_0$};
  \foreach \n in {1,...,5}{
   \node[v]at(v\panel\n){$\mathcal V_{\n}$};}
 \end{scope}
}
\node[explanation]at(6.22,8.25)
 {\textbf{(a) Add one node}\\[5pt]
  Start from $\mathcal E_{0,1}$.
  Each new node is linked to two earlier nodes.
  At generic positions, these distances prevent small relative motion.};
\node[explanation]at(6.22,5)
 {\textbf{(b) Remove a task edge}\\[5pt]
  For $\mathcal H=\mathcal C_5$,
  $\mathcal H-\mathcal E_{1,5}$ stays connected.
  Adding the reference edges gives generic rigidity as in (a).};
\node[explanation]at(6.22,1.75)
 {\textbf{(c) Remove a reference edge}\\[5pt]
  $\mathcal H-\mathcal V_1$ with its reference edges is generically rigid.
  Two task edges remain to constrain $\mathcal V_1$ after deleting
  $\mathcal E_{0,1}$.};
\end{tikzpicture}
\vspace*{-1em}
\caption{Adding and removing edges in the proof of Lemma~\ref{lem:cone}:
(a) building a rigid graph along a spanning tree, (b) deleting a task edge,
and (c) deleting a reference edge.}
\label{fig:rigidity-steps}
\end{figure}

Next, we show that adding all reference edges to a connected task graph
gives a generically rigid required graph~\cite{connelly2010coning}.
This construction will be used below to verify rigidity after
edge deletion. We choose a spanning tree of the task graph, which connects all its vertices with the minimum
number of edges. We start with $\mathcal V_0$ and one tree vertex joined
by their reference edge. The fixed distance allows only translation
and rotation of this pair, which forms the rigid starting graph.
We add the remaining tree vertices one at a time, choosing a vertex
whose tree neighbor is already present. In this way, the new vertex is connected to the existing tree neighbor
by a tree edge and to $\mathcal V_0$ by a reference edge. We fix the position and orientation of the rigid part.
The task distance and reference distance place the new vertex on two
circles centered at its existing tree neighbor and $\mathcal V_0$,
respectively. The new vertex lies at an intersection of these circles,
which prevents continuous motion while both distances remain fixed. Repeating this
construction establishes rigidity at generic positions. The task edges of $\mathcal H$ outside the selected
spanning tree impose additional distance constraints, so adding these
edges preserves rigidity of the full required graph.

In Fig.~\ref{fig:rigidity-steps}(a), the spanning tree is the path
through $\mathcal V_1,\cdots,\mathcal V_5$ in this order. The two highlighted edges
$\mathcal E_{4,5}$ and $\mathcal E_{0,5}$ connect the last added node
$\mathcal V_5$ to the existing node $\mathcal V_4$ and node $\mathcal V_0$.
Their distances prevent continuous motion of $\mathcal V_5$ relative
to the existing graph.

\subsection{Sufficiency and Rigidity after Edge Deletion}
To prove sufficiency, we suppose that $\mathcal H$ is 2-vertex-connected.
First, we verify that $\mathcal F$ remains connected when at most two
vertices are deleted.
If $\mathcal V_0$ remains, its reference edges connect all remaining
vertices. Otherwise, the remaining graph is $\mathcal H$ or
$\mathcal H-\mathcal V_n$, both connected. Then, we verify rigidity
after deleting either type of edge.

\textbf{Case 1: A task edge $\mathcal E\in\mathcal E_{\mathcal H}$.}
First, we show that deleting $\mathcal E$ leaves the task
graph connected, so the preceding rigidity construction remains
applicable. If removing $\mathcal E$ disconnected $\mathcal H$, this edge
would be the only connection between two parts of the task graph.
Deleting an endpoint in a part containing other nodes would then
disconnect those nodes from the other part. This contradicts the
requirement that $\mathcal H$ remain connected after any one node is
deleted. Thus, $\mathcal H-\mathcal E$ remains connected. Since all reference
edges remain, we can apply the preceding construction to show that
$\mathcal F-\mathcal E$ is generically rigid. In Fig.~\ref{fig:rigidity-steps}(b), deleting
the dashed edge $\mathcal E_{1,5}$ from the cycle graph $\mathcal C_5$
leaves the path through $\mathcal V_1,\cdots,\mathcal V_5$.
Together with the reference edges, the remaining path forms the rigid
graph constructed in panel (a).

\textbf{Case 2: A reference edge $\mathcal E_{0,n}$.}
First, we temporarily remove $\mathcal V_n$ and all edges joined
to this node to examine the graph after deleting $\mathcal E_{0,n}$. The remaining task graph $\mathcal H-\mathcal V_n$ is connected and,
together with $\mathcal V_0$ and the reference edges joining $\mathcal V_0$
to all vertices of $\mathcal H-\mathcal V_n$, forms a generically rigid graph
by the preceding spanning-tree construction.
Then, we restore $\mathcal V_n$ and its task edges, while leaving
$\mathcal E_{0,n}$ deleted. Since $\mathcal H$ is 2-vertex-connected,
$\mathcal V_n$ has at least two distinct task neighbors in the rigid
part. If $\mathcal V_n$ had only one task neighbor,
deleting that neighbor would disconnect $\mathcal V_n$ from the rest
of $\mathcal H$. The distances to two of these neighbors restrict $\mathcal V_n$
to the intersections of two circles, as above. Thus, $\mathcal V_n$
cannot move continuously relative to the rigid part, and
$\mathcal F-\mathcal E_{0,n}$ is generically rigid.

In Fig.~\ref{fig:rigidity-steps}(c), the deleted reference edge is
$\mathcal E_{0,1}$, shown dashed. The path through
$\mathcal V_2,\mathcal V_3,\mathcal V_4,\mathcal V_5$, together with
$\mathcal V_0$ and the reference edges joining $\mathcal V_0$ to these four
vertices, forms a rigid graph. The highlighted task edges $\mathcal E_{1,2}$ and
$\mathcal E_{1,5}$ connect $\mathcal V_1$ to two nodes of this rigid
graph, so their distances prevent continuous motion of $\mathcal V_1$
after $\mathcal E_{0,1}$ is removed.

Thus, $\mathcal F$ is 3-vertex-connected and remains generically rigid
after any edge deletion. The planar criterion gives generic
global rigidity. This proves sufficiency and
completes the proof.

\section{Proof of Theorem~\ref{thm:joint}}\label{app:joint}
First, we establish a lower bound on the joint minimum round
count. Then, we verify two graph and schedule constructions that achieve it. Every non-reference node transmits at least once, so
\begin{equation}
 M\le\sum_{n=1}^{M}w_n
   =\sum_{r=1}^{R}K_r\le RK_{\max}.
 \label{eq:app-load}
\end{equation}
By Lemma~\ref{lem:cone}, $\mathcal H$ is 2-vertex-connected and has
a task edge whose reciprocal coverage requires two different rounds.
Thus, $R\ge\max\{2,\lceil M/K_{\max}\rceil\}$. In the constructions
below, we set $R$ to this lower bound and assign each node one
transmission in round $c_n$.
The general construction applies when $R\ge3$ or $M$ is even.
The special construction adjusts the task graph when $R=2$ and
$M$ is odd, so the two-round lower bound remains achievable.

\subsection[Cycle Graph and Round Assignment in the General Case]{Cycle Graph and Round Assignment in the General Case}
\begin{figure}[t]
\centering
\begin{tikzpicture}[x=1cm,y=1cm,
 every node/.style={font=\fontsize{9}{11}\selectfont},
 v/.style={circle,draw,line width=.7pt,minimum size=9mm,inner sep=.2pt,align=center},
 c1/.style={v,draw=blue!70!black,fill=blue!18},
 c2/.style={v,draw=orange!80!black,fill=orange!22},
 c3/.style={v,draw=teal!70!black,fill=teal!18}]
\path[use as bounding box](0,0)rectangle(8.5,4.85);
\node[font=\fontsize{9}{11}\selectfont\bfseries]at(2.05,4.61)
 {(a) Before exchange};
\node[font=\fontsize{9}{11}\selectfont\bfseries]at(6.45,4.61)
 {(b) After exchange};
\begin{scope}[shift={(2.05,2.55)}]
 \foreach \n in {1,...,7}{
  \coordinate(a\n)at({90-360/7*(\n-1)}:1.4);}
 \draw[thick](a1)--(a2)--(a3)--(a4)--(a5)--(a6)--(a7);
 \draw[red!75!black,dashed,line width=1.2pt](a7)--(a1);
 \foreach \n/\colorid in {1/1,2/2,3/3,4/1,5/2,6/3,7/1}{
  \node[c\colorid]at(a\n){$\mathcal V_{\n}$};}
\end{scope}
\draw[-{Stealth[length=1.6mm]},thick](3.95,2.55)--(4.55,2.55);
\begin{scope}[shift={(6.45,2.55)}]
 \foreach \n in {1,...,7}{
  \coordinate(b\n)at({90-360/7*(\n-1)}:1.4);}
 \draw[thick](b1)--(b2)--(b3)--(b4)--(b5)--(b6)--(b7)--(b1);
 \foreach \n/\colorid in {1/1,2/2,3/3,4/1,5/2}{
  \node[c\colorid]at(b\n){$\mathcal V_{\n}$};}
 \node[c1]at(b6){$\mathcal V_6$};
 \node[c3]at(b7){$\mathcal V_7$};
\end{scope}
\foreach \g/\xx in {1/1.10,2/3.65,3/6.20}{
 \node[c\g,minimum size=3.5mm,inner sep=0pt]at(\xx,.35){};
 \node[anchor=west]at({\xx+.28},.35){Group \g};}
\end{tikzpicture}
\caption{Exchanging the round assignments of $\mathcal V_6$ and
$\mathcal V_7$ on $\mathcal C_7$ for $M=7$, $R=3$, and $K_{\max}=3$.
After the exchange, adjacent nodes belong to different groups.
Each group transmits in one round, and the group sizes remain 3, 2,
and 2. Thus, the numbers of transmitters in the 3 rounds are unchanged
and satisfy the receiver-capacity limit.}
\label{fig:balancing}
\end{figure}
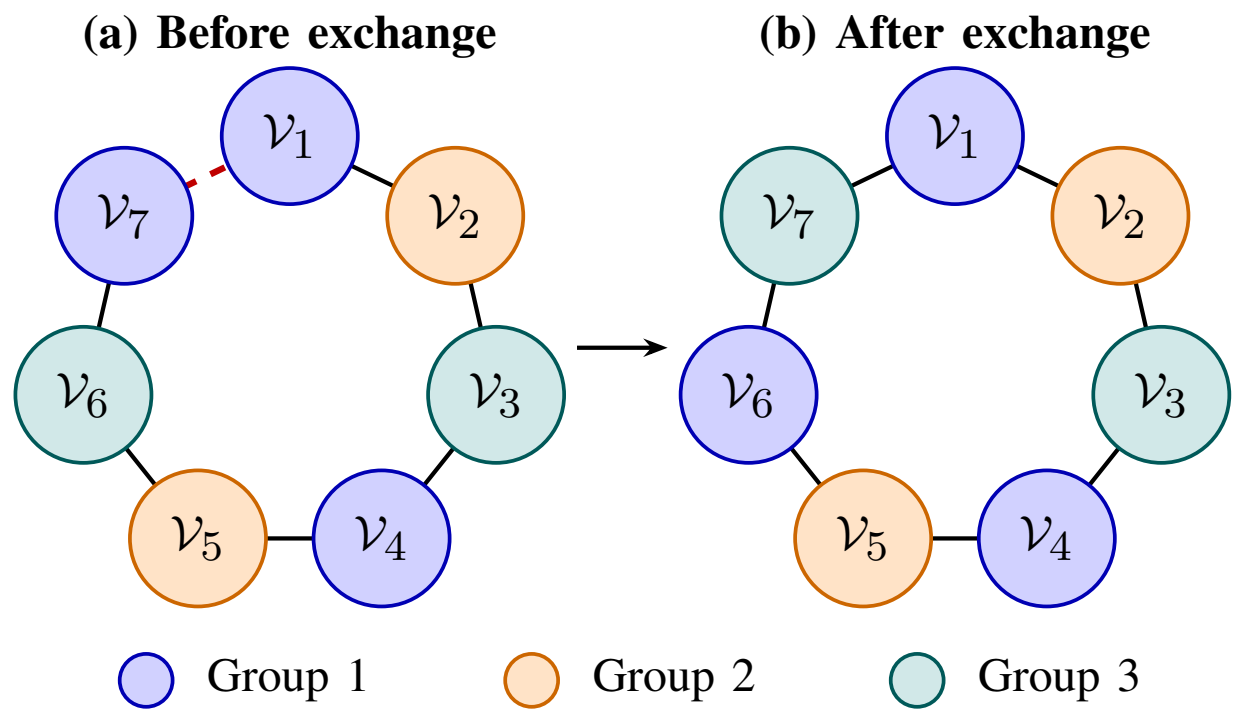

When $R\ge3$ or $M$ is even, we choose $\mathcal H=\mathcal C_M$ in the
order $\mathcal V_1,\cdots,\mathcal V_M$ and assign
$c_n=1+((n-1)\bmod R)$. Adjacent vertices have different round indices,
except that $\mathcal V_M$ and $\mathcal V_1$ share the same index
when $M\bmod R=1$. In that case, $R\ge3$, since $R=2$ is used here
only for even $M$.
When the endpoints of the closing edge share a round index, we exchange the round indices of the last two nodes,
after which we have $c_{M-1}=1$ and $c_M=R$, while $c_{M-2}=R-1$
remains unchanged.
The only affected task edges are
$\mathcal E_{M-2,M-1}$, $\mathcal E_{M-1,M}$, and $\mathcal E_{M,1}$.
After the exchange, their endpoint round indices are $(R-1,1)$,
$(1,R)$, and $(R,1)$, respectively. Since $R\ge3$, each of these edges
joins nodes assigned to different rounds. For all other task edges, the round indices of both
endpoints remain unchanged. The exchange preserves the round loads,
which differ by at most one under the repeated assignment. Thus, we have
\begingroup
\begin{equation}
 \begin{aligned}
 K_r&\in\left\{
       \left\lfloor\frac{M}{R}\right\rfloor,
       \left\lceil\frac{M}{R}\right\rceil
       \right\},\qquad 1\le r\le R,\\
 &\max\limits_{1\le r\le R}K_r
   =\left\lceil\frac{M}{R}\right\rceil\le K_{\max}.
 \end{aligned}
 \label{eq:app-balanced-load}
\end{equation}
\endgroup
Every task edge has reciprocal coverage, and deleting any vertex of
$\mathcal C_M$ leaves a connected path. Lemma~\ref{lem:cone}
guarantees generic global rigidity of $\mathcal F$.
In Fig.~\ref{fig:balancing}, we illustrate the exchange for $M=7$,
$R=3$, and $K_{\max}=3$. Initially, $\mathcal V_1$ and $\mathcal V_7$
both transmit in round 1, so the closing edge $\mathcal E_{1,7}$ lacks
reciprocal coverage. Exchanging $c_6$ and $c_7$ gives $c_6=1$ and $c_7=3$,
so the two endpoints of every task edge transmit in different rounds. The round loads remain $(3,2,2)$, which still satisfy the
receiver-capacity limit.

\subsection[Graph Adjustment in the Special Two-Round Case]{Graph Adjustment in the Special Two-Round Case}
When $R=2$ and $M$ is odd, alternating round indices
1 and 2 around $\mathcal C_M$ assigns the same index to the first and
last nodes. In this case, the closing task edge lacks reciprocal coverage.
To avoid this conflict, we modify the task graph. First, we use the even
cycle graph $\mathcal C_{M-1}$ and assign alternating round indices 1
and 2, starting with 1 at $\mathcal V_1$. Then, we assign node
$\mathcal V_M$ to round 2 and connect it to $\mathcal V_1$ and $\mathcal V_3$. These two distinct neighbors both
transmit in round 1, so every task edge has reciprocal coverage.
The round loads are $(M-1)/2$ and $(M+1)/2$, both at most $K_{\max}$
because $M\le2K_{\max}$. Deleting $\mathcal V_M$ leaves
$\mathcal C_{M-1}$. If a cycle vertex is deleted instead, the even cycle becomes a
path. At least one of the two task edges of $\mathcal V_M$ remains,
so this node stays connected to the path. Hence,
$\mathcal H$ is 2-vertex-connected, and Lemma~\ref{lem:cone} again
guarantees generic global rigidity. Both constructions achieve the
lower bound with one transmission per non-reference node, giving
$N+1$ individual broadcasts and $L=R+2$ rounds per ranging cycle.
This completes the proof.

\begin{remark}
For small networks with $N\le4$, the minimum intermediate-round counts
are $1,2,3$ for $N=2,3,4$, respectively. Specifically, $N=2$ needs one non-reference transmission.
For $N=3$, the required triangle needs the
2 non-reference nodes to transmit in separate rounds. For $N=4$,
generic planar global rigidity requires $\mathcal K_4$, since
generic global rigidity requires 3-vertex-connectivity~\cite{jackson2005}.
The subgraph of $\mathcal K_4$ induced by the non-reference nodes
is $\mathcal K_3$, which needs 3 rounds by
Theorem~\ref{thm:fixed}. All three counts can be achieved by sequential
transmission under any positive receiver capacity.
\end{remark}

\section{Proof of Proposition~\ref{prop:edges}}\label{app:edges}
First, we derive a lower bound on the number of required distances. Then, we identify when a two-round schedule requires an additional task edge.
By Lemma~\ref{lem:cone}, the task graph $\mathcal H$
remains connected after any one node is deleted. If a node had only one
task neighbor, deleting that neighbor would isolate the node. Hence,
every node in $\mathcal H$ has at least two task neighbors. Since each task edge contributes twice to the sum of node
degrees, adding the $M$ reference edges gives
\begingroup
\begin{equation}
 |\mathcal E_{\mathcal F}|
 =M+\frac{1}{2}\sum_{n=1}^{M}D_n
 \ge2M=2N-2.
 \label{eq:app-edge-count}
\end{equation}
\endgroup
The cycle graph
construction in Section~\ref{sec:closed-form-design} achieves this bound at the
minimum round count whenever $M$ is even or $R^\star_{\rm joint}\ge3$.

We now examine the transmit groups for odd $M$ with two rounds.
Each node needs to transmit and to receive from its task neighbors.
A node transmitting in both rounds would have no listening round.
Therefore, each node transmits in one of the two rounds
and listens in the other. The transmit support of each node is either
$\{1\}$ or $\{2\}$, and each task edge joins nodes with different supports.

If $|\mathcal E_{\mathcal H}|=M$, we show that the task edges
must form a single cycle. The degrees sum to $2M$ and are
all at least 2, so each node has exactly two task neighbors. Starting from any node, we follow task edges without
immediately returning along the edge just used. At each new node,
exactly one edge remains for continuing the path, so there is no branch.
Since the graph is finite and every node has exactly two neighbors,
the path eventually returns to the starting node and forms a cycle.
Each node on this cycle already uses both of its task edges within the
cycle and cannot connect to any node outside it. As $\mathcal H$ is
connected, the cycle contains all $M$ nodes.

For odd $M$, this cycle cannot satisfy the two-round assignment.
Alternating round indices 1 and 2 along this cycle assigns the same index to the
first and last nodes. The closing task edge has no reciprocal
coverage, contradicting the required schedule. Thus, we have
$|\mathcal E_{\mathcal H}|\ge M+1$, which gives
$|\mathcal E_{\mathcal F}|\ge2N-1$. The two-round construction in
Section~\ref{sec:closed-form-design} has exactly $M+1$ task edges and $M$ reference
edges, which achieves this bound. This completes the proof.

\bibliographystyle{IEEEtran}
\bibliography{references}
\begingroup
\makeatletter
\def\@IEEEBIOskipN{2\baselineskip}
\makeatother
\begin{IEEEbiography}[{\includegraphics[width=1in,height=1.25in,trim=72.74091bp 102.73716bp 75.85302bp 5.99925bp,clip,keepaspectratio]{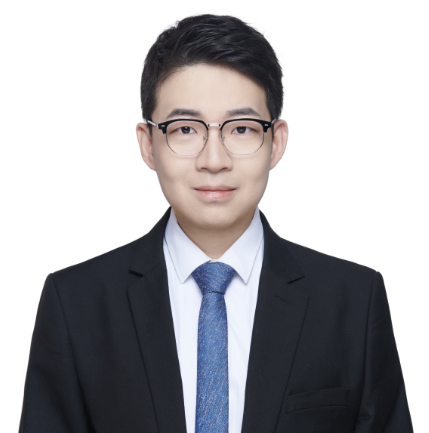}}]{Zijian Zhang}
	(Member, IEEE) received the B.E. degree and Ph.D. degree in electronic engineering from Tsinghua University, Beijing, China, in 2020 and 2025, respectively. His research area is network localization and beyond massive MIMO for future wireless systems. He has received Tsinghua Science Technology Best Paper Award in 2024 and the IEEE Signal Processing Society Young Author Best Paper Award in 2025. In 2024, he won the Special Scholarship of Tsinghua University, which annually awards 10 students out of over 40,000 graduate students in Tsinghua University. He was listed in Stanford World's Top 2\% Scientists in 2023 and 2025.
\end{IEEEbiography}

\begin{IEEEbiography}
    [{\includegraphics[width=1in,height=1.25in,clip,keepaspectratio]{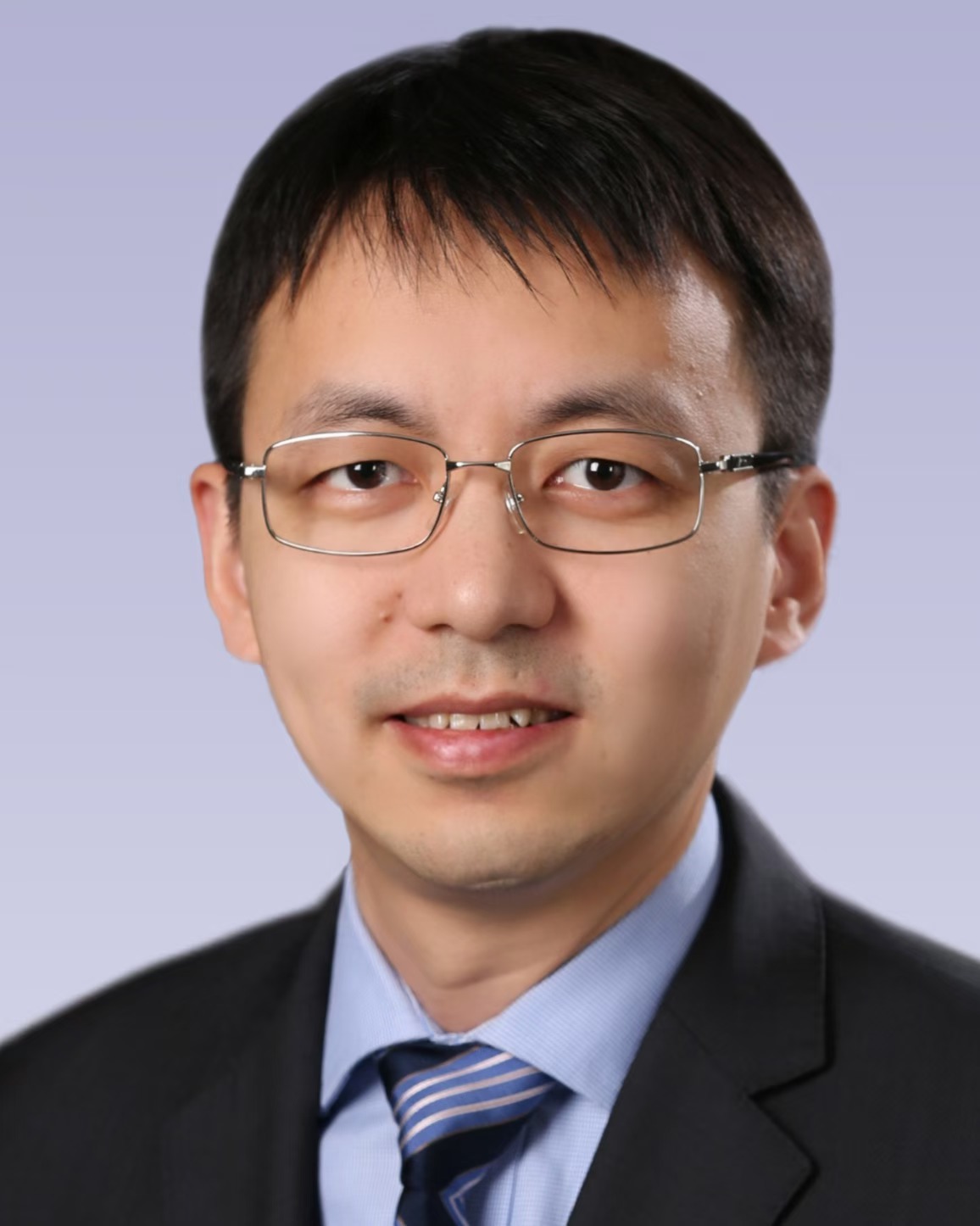}}]
{Yuan Shen} (Senior Member, IEEE) received the B.E. degree in electronic engineering from Tsinghua University in 2005, and the S.M. and Ph.D. degrees in electrical engineering and computer science from the Massachusetts Institute of Technology (MIT) in 2008 and 2014, respectively. He is currently a Full Professor with the Department of Electronic Engineering, Tsinghua University. His research interests include integrated sensing and communication, multi-agent systems and AI for science. His papers have received the IEEE ComSoc Fred W. Ellersick Prize and several best paper awards from IEEE conferences. He has served as the TPC Symposium Co-Chair for IEEE ICC and IEEE Globecom for several times. He was the Elected Chair for the IEEE ComSoc Radio Communications Committee from 2019 to 2020. He is currently an Editor of the IEEE \textsc{TRANSACTIONS ON COMMUNICATIONS}, IEEE \textsc{TRANSACTIONS ON WIRELESS COMMUNICATIONS}, and \textsc{China Communications}.
\end{IEEEbiography}

\endgroup
\end{document}